\documentclass[12pt, reqno]{amsart}
\allowdisplaybreaks

\usepackage[all]{xy}
\usepackage{amssymb} 

\newcommand{\morespace}{}

\newcommand{\wt}{\widetilde}
\newcommand{\wh}{\widehat}
\newcommand{\ov}{\overline}
\newcommand{\os}[2]{\overset{#1}{#2}}

\newcommand{\pnt}[1]{\lbrace #1 \rbrace}

\newtheorem{thm}{Theorem}
\newtheorem{prop}{Proposition}

\begin{document}
\morespace

\title[Orthoalgebras without Bivaluations]{New Families of Finite Coherent Orthoalgebras 
without Bivaluations}

\author[A. E. Ruuge]{Artur E. Ruuge}

\author[F. van Oystaeyen]{Freddy van Oystaeyen}


\begin{abstract}
\morespace
In the present paper we study the following problem: 
how to construct a coherent orthoalgebra which has only a finite number of elements, 
but at the same time does not admit a bivaluation 
(i.e. a morphism with a codomain being an orthoalgebra with just two elements). 
This problem is important in the perspective of Bell-Kochen-Specker theory, since 
one can associate such an orthoalgebra
to every saturated non-colorable finite configuration of projective lines. 
The first result obtained in this paper provides   
a general method for constructing finite orthoalgebras. 
This method is then applied to obtain a new infinite family of finite coherent 
orthoalgebras that do not admit bivaluations. 
The corresponding proof is combinatorial and yields a description of the groups of symmetries 
for these orthoalgebras. 

\end{abstract}

\maketitle

\vspace{1 true cm}

\section{Introduction}

George W. Mackey  
formulated 
in his book \cite{Mackey}
the axiomatics of non-relativistic quantum mechanics 
based on the notion of an \emph{orthomodular poset}. 
That is just a partially ordered set equipped with an involution, such that 
certain axioms hold. These axioms are chosen such that 
the elements of this poset may be identified with binary observables of a quantum system. 
Compared to the traditional axiomatics in terms of linear operators 
on Hilbert spaces \cite{Neumann}, this system focuses on the \emph{logical} 
aspects of quantum theory. 
In fact, the Hilbert space is introduced only at the final stage in a completely 
\emph{ad hoc} manner. 

In alternative terminology, an orthomodular poset is called a \emph{coherent orthoalgebra}, and 
an orthoalgebra is a particular case of an \emph{effect algebra}. 
Let us provide some motivation for the introduction of these notions. 
Consider a Hilbert space $\mathcal{H}$ over $\mathbb{C}$, and denote by $L (\mathcal{H})$ 
a collection of closed linear manifolds in it. For every $U \in L (\mathcal{H})$, we have 
an orthogonal projector $\wh{\pi}_{U}$ on $U$, 
which represents an observable with two possible values, $0$ and $1$. 
Two observables represented by $\wh{\pi}_{U}$ and $\wh{\pi}_{U_1}$, $U, U_1 \in L (\mathcal{H})$, 
are \emph{compatible} iff their commutator $[\wh{\pi}_{U}, \wh{\pi}_{U_1}] = 0$. 
The first step towards the notion of an effect algebra is based on the following remark.  
The mentioned commutator vanishes iff  $\mathcal{H}$ splits into an orthogonal sum 
$\mathcal{H} = Z \oplus V \oplus V_1 \oplus W$, such that $Z \oplus V = U$ and $Z \oplus V_1 = U_1$. 
The idea is to reformulate everything in terms of orthogonal decomposition.

Consider $\cdot \oplus \cdot$ as a partially defined binary operation on $L (\mathcal{H})$ with 
domain of definition consisting of all pairs $(U, U_1)$ such that $U_1 \subset U^{\perp}$. 
Note that $U_1 \subset U^{\perp}$ is equivalent to $U \subset U_{1}^{\perp}$.  
Consider $L (\mathcal{H})$ as a partially ordered set with respect to inclusion $\subset$. 
Then the map $U \mapsto U^{\perp}$ is an involution on $L (\mathcal{H})$, 
since $U^{\perp \perp} = U$ and for all $U$ and $U_1$ we have  
$U \subset U_1 \Leftrightarrow U^{\perp} \supset U_{1}^{\perp}$. 
Note, that it is possible to express the partial order $\subset$ in terms of 
the $\cdot \oplus \cdot$ operation: 
$U \subset U_1$ iff $\exists V : V \oplus U = U_1$. 
The involution $(\cdot)^{\perp} : L (\mathcal{H}) \to L (\mathcal{H})$, 
admits a similar characterization. 
For every $U$, there exists a unique $U_{1}$, such that $U_{1} \oplus U = \mathcal{H}$; this $U_{1}$ is 
precisely $U^{\perp}$. 

Take any $U, U_1 \in L (\mathcal{H})$. If the corresponding two observables are compatible,  
then the following formulae are valid: 
\begin{equation}
\begin{gathered} 
\inf \lbrace U, U_1 \rbrace \oplus U^{\perp} = 
U_1 \oplus \sup \lbrace U, U_{1} \rbrace^{\perp}, \\
\inf \lbrace U, U_1 \rbrace \oplus U_{1}^{\perp} = 
U \oplus \sup \lbrace U, U_{1} \rbrace^{\perp}. 
\end{gathered}
\label{uu1_compatible}
\end{equation}
Moreover, it is not difficult to prove that 
$\wh{\pi}_U$ and $\wh{\pi}_{U_1}$ are compatible exactly when these two equalities 
\eqref{uu1_compatible} are valid. 
Which properties of $\cdot \oplus \cdot$ are actually needed in this proof?  
It turns out that 
it is convenient to capture these properties within the notion of an \emph{effect algebra}.

Let $S$ be a set, and $R \subset S \times S$ -- a relation on $S$. 
Let $\cdot \oplus \cdot : R \to S$, $(x, y) \mapsto x \oplus y$, be a map.  
Let $\mathbf{0}$ and $\mathbf{1}$ be two elements in $S$, such that $\mathbf{1} \not = \mathbf{0}$.  
The algebraic structure $(S, \oplus, \mathbf{0}, \mathbf{1})$ is called an \emph{effect algebra} 
if for all $x, y, z \in S$ the following conditions are satisfied:

\begin{itemize}
\item[1)] if $x \oplus y$ is defined, then $y \oplus x$ is defined and 
$y \oplus x = x \oplus y$; 
\item[2)] if $(x \oplus y) \oplus z$ is defined, then $x \oplus (y \oplus z)$ is defined 
and $x \oplus (y \oplus z) = (x \oplus y) \oplus z$; 
\item[3)] $x \oplus \mathbf{0} = x$; 
\item[4)] if $x \oplus y = x \oplus z$, then $y = z$; 
\item[5)] there exists $x^* \in S$, such that $x^{*} \oplus x = \mathbf{1}$.
\item[6)] if $x \oplus \mathbf{1}$ is defined, then $x = \mathbf{0}$;  
\end{itemize}

Note that for each $x$, the element $x^{*}$ is uniquely defined. 
Hence, to every effect algebra $X = (S, \oplus, \mathbf{0}, \mathbf{1})$ one associates a map 
$(\cdot)^{*} : S \to S$, $x \mapsto x^{*}$. 
The set $S$ is termed the \emph{ground} set of $X$. 
  
An effect algebra is called an \emph{orthoalgebra}, if for any element $x$ of the ground set, 
such that $x \oplus x$ is defined, we have $x = \mathbf{0}$. 
Note that this property together with the first five axioms, implies the sixth axiom. 
An othoalgebra is called \emph{coherent} if 
for all $x$, $y$, and $z$ in the ground set, such that 
$x \oplus y$, $y \oplus z$, and $z \oplus x$ 
are defined, the $x \oplus y \oplus z$ is defined.

The basic example of an effect algebra is, of course, the following: 
$S = L (\mathcal{H})$, 
$\oplus$ -- the orthogonal sum defined for all $(U, U_1)$ such that $U_1 \subset U^{\perp}$, 
$\mathbf{0} = \theta_{\mathcal{H}}$ -- the trivial subspace of $\mathcal{H}$, and 
$\mathbf{1} = \mathcal{H}$.  
Denote this effect algebra by $\mathbb{L} (\mathcal{H})$. 
In fact, it is a coherent orthoalgebra. 
Just as for $\mathbb{L} (\mathcal{H})$, 
one can define for every effect algebra $X = (S, \oplus, \mathbf{0}, \mathbf{1})$ a 
partial order $\preccurlyeq$ on the ground set $S$ (termed the \emph{standard} partial order): 
$\forall x, y \in S: x \preccurlyeq y : \Leftrightarrow \exists x_1 : x_1 \oplus x = y$.  
The map $(\cdot)^{*}$ is an involution with respect to $\preccurlyeq$. 
It is possible to imitate the notion of compatibility on any effect algebra as follows: 
call two elements $U, U_1 \in S$ \emph{compatible}, if the set $\lbrace U, U_1 \rbrace$ 
has infimum and supremum (with respect to the standard partial order), 
and the formulae of the form \eqref{uu1_compatible} (with $\perp$ 
replaced by $\ast$) are valid. 
Such a definition of compatibility, is additionally justified by the following fact: 
for any compatible $U$ and $U_1$, there exists 
a decomposition of $\mathbf{1}$ of the form $\mathbf{1} = Z \oplus V \oplus V_1 \oplus W$, 
such that $Z \oplus V = U$ and $Z \oplus V_1 = U_1$. 

Since the notion of a coherent orthoalgebra captures up to certain extent the essential properties of 
$L (\mathcal{H})$, it presents special interest to investigate 
the case when the ground set is \emph{finite}. By that one may try to imitate quantum 
mechanics on a finite set. The latter is not only conceptually interesting, but also 
can be important for the computational methods.  
Of course, it is necessary to have a ``complicated enough'' example for this case. 

It is natural to introduce a category of effect algebras $\mathcal{E}$ with morphisms 
$f : (S, \oplus, \mathbf{0}, \mathbf{1}) \to (S', \oplus', \mathbf{0}', \mathbf{1}')$ 
being the maps $\bar{f} : S \to S'$ such that $\bar{f} (\mathbf{0}) = \mathbf{0}'$, 
$\bar{f} (\mathbf{1}) = \mathbf{1}'$, and $\bar{f} (x \oplus y) = \bar{f} (x) \oplus' \bar{f} (y)$, 
whenever $x \oplus y$ is defined. 
The composition of morphisms is defined by the composition of the corresponding maps. 
Consider the most simple effect algebra that can be -- the effect algebra with only two elements -- 
$\mathbf{0}$ and $\mathbf{1}$. 
This is an initial object in the category of effect algebras.  
There is only one way to define $\oplus$ in this case: 
$\mathbf{0} \oplus \mathbf{0} := \mathbf{0}$, 
$\mathbf{0} \oplus \mathbf{1} = \mathbf{1} \oplus \mathbf{0} := \mathbf{1}$, 
and $\mathbf{1} \oplus \mathbf{1}$ -- undefined. 
Denote this object by $\mathbb{B}$ and call it the \emph{minimal Boolean} effect algebra. 
The other example of an effect algebra that has been described above is $\mathbb{L} (\mathcal{H})$. 
Call it the \emph{Hilbert} effect algebra. 
Is it possible to have an arrow from  $\mathbb{L} (\mathcal{H})$ to $\mathbb{B}$  
in the category $\mathcal{E}$?
The answer is well known from functional analysis (Gleason's theorem) and is negative.  
At the same time there is another important example of an effect algebra $(S, \oplus, \mathbf{0}, \mathbf{1})$, 
for which such an arrow exists.  
Let $S = \mathcal{F}$, where $\mathcal{F}$ is some $\sigma$-algebra of subsets of a set $\Omega$.  
Define $U \oplus U_1$ as $U \cup U_1$ for all \emph{disjoint} $U, U_1 \in \mathcal{F}$. 
Put $\mathbf{0} = \emptyset$ and $\mathbf{1} = \Omega$. This defines an effect algebra, 
denoted by $\mathbb{W} (\mathcal{F})$ and called \emph{Kolmogorov} effect algebra. 
Any $\mathbb{W} (\mathcal{F})$ admits a morphism $f$ to $\mathbb{B}$: 
one may fix any $\omega \in \Omega$ and for each $U \in \mathcal{F}$ put 
$\bar f (U) = \mathbf{1}$ if $U \ni \omega$, and $\mathbf{0}$ -- otherwise. 

The Kolmogorov and Hilbert effect algebras, 
$\mathbb{W} (\mathcal{F})$ and $\mathbb{L} (\mathcal{H})$, 
are different, and this is clear 
if one looks at all morphisms ending in the minimal Boolean effect algebra $\mathbb{B}$.  
This motivates the following mathematical problem. 
For any $X \in \mathcal{E}$, let us call an arrow $f : X \to \mathbb{B}$ (if it exists) 
a \emph{bivaluation}. 
Denote by $\textit{for}$ the forgetful functor from $\mathcal{E}$ to the category of sets, 
$\textit{for} : \mathcal{E} \to \mathbf{Sets}$. 
One is required to find in $\mathcal{E}$ such objects $X$, 
which do not admit a bivaluation, but have a finite ground set $\emph{for} (X)$. 
In the present paper an infinite family of such objects is constructed.    

Let us make several bibliographical remarks to conclude the introduction. 
The analysis of logical foundations of quantum mechanics has been initiated in the famous paper 
by G. Birkhoff and J. von Neumann \cite{BirkhoffNeumann}. 
The new wave of interest to this subject is motivated by the recent developments in quantum computing 
technology. For an up to date discussion of effect algebras, orthoalgebras, 
and similar structures, one should refer to the monograph \cite{DvurPulm}. 
The terms `effect algebra' and `orthoalgebra' 
were suggested in \cite{FoulisBennett} and 
\cite{FoulisGreechieRuttimann}, respectively. 
The importance of orthoalgebras is also clear in the perspective of the consistent 
histories approach to quantum theory \cite{Isham}. 

The results obtained in the present paper 
are related to the results of \cite{Ruuge}, \cite{RuugeFVO}, and 
may be viewed as their generalization. 
The orthoalgebras described below yield a family of `indeterministic objects' 
in the terminology of \cite{Ruuge}. 
Every saturated (in the sense of \cite{RuugeFVO}) 
Kochen-Specker-type configuration of projective lines 
naturally yields a finite orthoalgebra not admitting a bivaluation.

\section{General construction}

How to construct a \emph{finite} orthoalgebra, which will look ``similar'' to the 
Hilbert orthoalgebra? The starting point can be the following. 
Consider a Hilbert space $\mathcal{H}$ over $\mathbb{C}$ of finite dimension $d$. 
Let $\mathbb{P} (\mathcal{H})$ denote the set of projective lines in $\mathcal{H}$. 
Consider the set $\mathcal{P}_{\perp} (\mathbb{P} (\mathcal{H}))$ 
consisting of all subsets 
$U \subset \mathbb{P} (\mathcal{H})$ satisfying the condition 
$\forall l, l_1 \in U : l_1 \not = l \Rightarrow l \perp l_1$. 
Note, that the empty set and any subset with only one element, belong to 
$\mathcal{P}_{\perp} (\mathbb{P} (\mathcal{H}) )$. 
There is a natural equivalence relation $\sim$ on this set: 
$U \sim U_1 :\Leftrightarrow \mathrm{span} U_1 = \mathrm{span} U$ 
(the span of the empty set is $\theta_{\mathcal{H}}$ by definition). 
It is clear, that the set $\mathcal{L} (\mathcal{H}) := \mathcal{P}_{\perp} (\mathbb{P} (\mathcal{H})) / \sim$ 
is in natural bijection with $L (\mathcal{H})$. Hence, the structure of orthoalgebra on $L (\mathcal{H})$ 
induces a structure of orthoalgebra on $\mathcal{L} (\mathcal{H})$. 
For $[U], [U_1] \in \mathcal{L} (\mathcal{H})$ ($[\cdot]$ denotes the equivalence class with respect to $\sim$), 
the value of $[U] \oplus [U_1]$ is defined iff $U \cap U_1 = \emptyset$ and 
$U \cup U_1 \in \mathcal{P}_{\perp} (\mathbb{P} (\mathcal{H}))$, and it is equal to $[U \cup U_1]$.    

This leads to the first (naive) idea of how to construct examples of finite orthoalgebras. 
Take a \emph{finite} set $A$ equipped with some relation $T \subset A \times A$, 
which is thought to imitate the orthogonality relation $\perp$. 
In analogy with $\mathcal{L} (\mathcal{H})$, consider the set  
\begin{equation*} 
\mathcal{P}_{T} (A) := \lbrace U \subset A \, | \, 
\forall l, l_1 \in U : l_1 \not = l \Rightarrow (l, l_1) \in T \rbrace, 
\end{equation*}
and try to find an  
equivalence relation $\sim$ on it, such that the formula $[U] \oplus [U_1] := [U \cup U_1]$ 
yields the structure of an orthoalgebra. It is necessary to describe this equivalence 
relation in terms of $T$. 
After that one faces the difficulty to 
find some reasonable conditions on $T$, entailing  
the axioms of an effect algebra. 

It turns out that there is a better idea. 
For \emph{any} $B \subset A$, denote 
\begin{equation*}
B^{T} := \lbrace l \in A \, | \, \forall l_1 \in A : 
l_1 \in B \Rightarrow (l_1, l) \in T \rbrace.
\end{equation*} 
Consider a map $\tau : \mathcal{P}_{T} (A) \to \mathcal{P} (A)$, $U \mapsto U^{T}$, and   
look at the image of this map, 
\begin{equation}
\mathcal{P}^{T} (A) := 
\mathrm{Im} \big( \mathcal{P}_{T} (A) \ni U \mapsto U^{T} \big). 
\label{PT_image}
\end{equation}
Take it as a ground set for the future orthoalgebra. 
Note, that if one specializes $A$ to $\mathbb{P} (\mathcal{H})$, and $T$ to the orthogonality 
relation $\perp$, then for $U, U_1 \in \mathcal{P}_{T} (A)$ one 
has $\tau (U) = \tau (U_1)$, whenever $\mathrm{span} U_1 = \mathrm{span} U$. 
It is natural to try to define the $\oplus$ operation by the formula 
\begin{equation} 
Q \oplus Q_1 := (Q \cup Q_{1})^{TT}, 
\label{oplus_constr}
\end{equation}   
for all $Q, Q_1 \in \mathcal{P}^{T} (A)$, such that 
$Q_{1} \subset Q^{T}$. 
Of course, it is necessary to impose some conditions on $T$, which ensure 
that $\oplus$ is well-defined, since  
the right-hand side is not \emph{a priori} in $\mathcal{P}^{T} (A)$. 
The axioms of an orthoalgebra will induce the other conditions on $T$. 

First, since $T$ is supposed to imitate the orthogonality relation $\perp$, one needs to 
require for all $l, l_1 \in A$, $l_1 \not = l$, the following:  
\begin{gather} 
(l, l_1) \in T \Leftrightarrow (l_1, l) \in T, 
\label{T_symm}
\\
(l, l) \not \in T. 
\label{T_nondiag}
\end{gather}
Impose one more condition: 
\begin{equation} 
\forall M \in \mathrm{Max} (\mathcal{P}_{T} (A); \subset) \, 
\forall B \subset M : B^{T} = (M \backslash B)^{TT}, 
\label{T_main}
\end{equation} 
where $\mathrm{Max} ( - )$ means taking the \emph{set} of all maximal subsets of the partially ordered set. 
Note that this condition is valid for the case $A = \mathbb{P} (\mathcal{H})$ and $T = \perp$. 
Let us say that $T$ is \emph{saturated} if it satisfies \eqref{T_main}.

\begin{thm}
Let $A$ be a finite non-empty set and $T$ -- a relation on $A$.  
Let $\mathcal{P}^{T} (A)$ be defined by \eqref{PT_image}. 
If $T$ satisfies the three conditions 
\eqref{T_symm}, \eqref{T_nondiag}, \eqref{T_main}, then 

1) $A$ and $\emptyset$ belong to $\mathcal{P}^{T} (A)$; 

2) $\oplus$ is well-defined by the formula \eqref{oplus_constr}; 

3) $(\mathcal{P}^{T} (A), \oplus, \emptyset, A)$ is a coherent orthoalgebra.  
\end{thm}

\noindent
\emph{Proof}. 
1) Since $\emptyset \in \mathcal{P}_{T} (A)$, and $\emptyset^{T} = A$, one has $A \in \mathcal{P}^{T} (A)$. 
Now, take any $M \in \mathrm{Max} (\mathcal{P}_{T} (A); \subset)$, put $B = M$, and apply the third 
condition on $T$ above. This yields: $M^{T} = (M \backslash M)^{TT} = \emptyset^{TT} = A^{T}$. 
If $A^{T}$ is not empty, then one can take any $l \in A^{T}$ and applying the definition 
of $(\cdot)^{T}$ claim, that $(l, l) \in T$. But this contradicts the first condition on $T$ above. 
Hence, $A^{T} = \emptyset$. Therefore, $\emptyset = M^{T} \in \mathcal{P}^{T} (A)$. 

2) Take any $Q, Q_1 \in \mathcal{P}_{T} (A)$, such that $Q_1 \subset Q^{T}$. 
It is necessary to show that $(Q \cup Q_{1})^{TT} \in \mathcal{P}^{T} (A)$. 
Invoking the main condition on $T$, represent $Q$ and $Q_1$ in the form 
$Q = U^T = (M \backslash U)^{TT}$, and $Q_1 = U_{1}^{T} = (M_{1} \backslash U_1)^{TT}$, 
where $U, U_1 \in \mathcal{P}_{T} (A)$, 
$M, M_1 \in \mathrm{Max} (\mathcal{P}_{T} (A); \subset)$, and 
$M \supset U$, $M_1 \supset U_1$. 
Since $Q_1 \subset Q^{T}$, 
for any $l \in (M \backslash U)^{TT}$ and any $l_1 \in (M_{1} \backslash U_1)^{TT}$, 
one has $(l, l_1) \in T$. 
Note, that due to the symmetry of $T$, for all $B \subset A$ there is an inclusion 
$B^{TT} \supset B$. 
Indeed, take any $\lambda_0 \in B$. In order to show, that $\lambda_0 \in B^{TT}$, 
one must show that $\forall \lambda_1 \in B^{T} : (\lambda_1, \lambda_0) \in T$. 
But the definition of $B^{T}$ implies that:  
$\forall \lambda \in B \, \forall \lambda_1 \in B^{T} : (\lambda, \lambda_1) \in T$. 
Since $T$ is symmetric, 
the order of appearance of $\lambda$ and $\lambda_1$ in $(\lambda, \lambda_1) \in T$ is unimportant, 
and one obtains $B \subset B^{TT}$. 
Now, return to $Q$ and $Q_1$.  
One has: $Q = U^{T} = (M \backslash U)^{TT}$, and 
$Q_1 = U_{1}^{T} = (M_{1} \backslash U_{1})^{TT}$. 
Take any $l \in M \backslash U$, and any $l_1 \in M_{1} \backslash U_{1}$. 
Since $M \backslash U \subset (M \backslash U)^{TT}$, and 
$M_{1} \backslash U_{1} \subset (M_{1} \backslash U_{1})^{TT}$, 
the elements $l$ and $l_1$ are in $Q$ and $Q_1$ respectively. 
From $Q_1 \subset Q^{T}$, one obtains $(l, l_1) \in T$. 
Therefore, $(M \backslash U) \cup (M_{1} \backslash U_{1}) \in \mathcal{P}_{T} (A)$. 
Now note, that 
for any $B, B_1 \subset A$, the definition of $(\cdot)^{T}$,  
without any assumptions on $T$, implies $(B \cup B_1)^{T} = B^{T} \cap B_{1}^{T}$. 
This together with the main condition, yields: 
\begin{multline*} 
(Q \cup Q_1)^{TT} = 
(Q^{T} \cap Q_{1}^{T})^{T} = 
(U^{TT} \cap U_{1}^{TT})^{T} = 
\\ 
= ((M \backslash U)^{T} \cap (M_{1} \backslash U_{1})^{T})^{T} = 
((M \backslash U) \cup (M_{1} \backslash U_{1}))^{TT}.
\end{multline*}

Before proceeding further, let us prove two simple auxiliary facts. 
Recall, that $A, \emptyset \in \mathcal{P}^{T} (A)$, and we have: 
$\emptyset^{T} = A$, $A^{T} = \emptyset$. Therefore $\emptyset^{TT} = \emptyset$, $A^{TT} = A$. 
Let us show that for any $Q_0 \in \mathcal{P}^{T} (A)$, the element $Q_0^{T} \in \mathcal{P}^{T} (A)$, and 
$Q_0^{TT} = Q_0$. Indeed, take any $Q_0$ and represent it in the form $Q_0 = U_0^{T}$, $U_0 \in \mathcal{P}_{T} (A)$. 
For any $M_0 \supset U_0$, $M_0 \in \mathrm{Max} (\mathcal{P}_{T} (A); \subset)$, the main condition implies: 
$Q_0^{T} = U_0^{TT} = (M_0 \backslash U_0)^{T}$. 
Since $M_0 \backslash U_0 \in \mathcal{P}_{T} (A)$, one has $Q_0^{T} \in \mathcal{P}^{T} (A)$. 
Now, for $Q_0^{TT}$, we have: 
$Q_0^{TT} = (U_0^{TT})^{T} = ((M_0 \backslash U_0)^{T})^{T} = 
(M_0 \backslash U_0)^{TT} = U_0^{T} = Q_0$ 
(we have used the main condition once more). 

Specializing $Q_0 \in \mathcal{P}^{T} (A)$ to $((M \backslash U) \cup (M_1 \backslash U_1))^{T}
= (M \backslash U)^{T} \cap (M_1 \backslash U_1)^{T} = U^{TT} \cap U_{1}^{TT} = 
(U^{T} \cup U_{1}^{T})^{T} = (Q \cup Q_{1})^{T}$, we obtain 
$\mathcal{P}^{T} (A) \ni Q_{0}^{T} = 
(Q \cup Q_1)^{TT}$, i.e. $\oplus$ is well defined. 

3) Let us start with the axioms of an effect algebra. Consider the first axiom.  
Take any $Q, Q_1 \in \mathcal{P}^{T} (A)$ such that $Q_1 \subset Q^{T}$. 
The latter inclusion means, that for any $l \in Q$ and any $l_1 \in Q_1$, the pair  
$(l, l_1) \in T$. Since $T$ is symmetric, $(l_1, l) \in T$. 
Hence, $Q_1 \subset Q^{T}$ is equivalent to $Q \subset Q_{1}^{T}$, 
i.e. $Q \oplus Q_1$ is defined iff $Q_1 \oplus Q$ is defined. 
We have: $Q \oplus Q_1 = (Q \cup Q_1)^{TT} = (Q_1 \cup Q)^{TT} = Q_1 \oplus Q$.

Next, let us verify the second axiom. 
Take any $Q, Q_1, Q_2 \in \mathcal{P}^{T} (A)$, and assume that 
$(Q \oplus Q_1) \oplus Q_2$ is defined. We have 
\begin{multline*} 
(Q \oplus Q_1) \oplus Q_2 = 
\big( (Q \cup Q_1)^{TT} \cup Q_2 \big)^{TT} = 
\big( (Q \cup Q_1)^{TTT} \cap Q_{2}^{T} \big)^{T} = \\ = 
\big( (Q \cup Q_1)^{T} \cap Q_{2}^{T} \big)^{T} = 
\big( Q^{T} \cap Q_{1}^{T} \cap Q_{2}^{T} \big)^{T} = 
\big( Q \cup Q_{1} \cup Q_{2} \big)^{TT}. 
\end{multline*}
Hence, if we can prove that $Q \oplus (Q_1 \oplus Q_2)$ is defined as well, 
then $Q \oplus (Q_1 \oplus Q_2) = (Q_1 \oplus Q_2) \oplus Q = 
\big( Q_{1} \cup Q_{2} \cup Q \big)^{TT}$, and then the second axiom is established. 
So, we assume $Q_1 \subset Q^{T}$ and $Q_2 \subset (Q \oplus Q_1)^{T}$, and need to verify two inclusions:  
$Q_2 \subset Q_{1}^{T}$ and $Q \subset (Q_1 \oplus Q_2)^{T}$. 
Recall that whenever $Q \oplus Q_1$ is defined, we know that 
$(Q \cup Q_1)^{T} \in \mathcal{P}^{T} (A)$.  
Derive: $Q_2 \subset (Q \oplus Q_1)^{T} = ((Q \cup Q_1)^{TT})^{T} = (Q \cup Q_{1})^{T} = Q^{T} \cap Q_{1}^{T}$. 
In particular $Q^{T} \cap Q_{1}^{T} \subset Q_{1}^{T}$ and therefore $Q_2 \subset Q_{1}^{T}$, i.e. 
the first inclusion is valid, i.e. $Q_1 \oplus Q_2$ is defined. 
Note, that we also have $Q_{2} \subset Q^{T}$, or, what is the same, $Q \subset Q_{2}^{T}$. 
Now, invoke the assumption $Q_1 \subset Q^{T}$, or, equivalently, $Q \subset Q_{1}^{T}$. 
Combining this with the previous fact, we obtain 
$Q \subset Q_{1}^{T} \cap Q_{2}^{T} = (Q_{1} \cup Q_{2})^{T} = 
(Q_{1} \cup Q_{2})^{TTT} = (Q_{1} \oplus Q_{2})^{T}$. 
Hence, the second inclusion is valid and by that the second axiom is established. 

Consider the third axiom. The candidate for $\mathbf{0}$ is $\emptyset$. 
For any $Q \in \mathcal{P}^{T} (A)$, $Q \oplus \mathbf{0}$ is defined, since $\emptyset \subset Q^{T}$. 
We have $Q \oplus \mathbf{0} = (Q \cup \emptyset)^{TT} = Q^{TT} = Q$. 
The third axiom is established. 

Before considering the fourth axiom, let us prove another general auxiliary fact. 
We know, that any $Q \in \mathcal{P}^{T} (A)$ can be represented in the form $Q = V^{TT}$, where 
$V \in \mathcal{P}_{T} (A)$ (take $U \in \mathcal{P}_{T} (A)$ and 
$M \in \mathrm{Max} (\mathcal{P}_{T} (A); \subset)$ such that 
$M \supset U$, and put $V = M \backslash U$). 
The element $Q^{T}$ is in $\mathcal{P}^{T} (A)$ as well. 
Hence, $Q^{T} = W^{TT}$, for some $W \in \mathcal{P}_{T} (A)$. 
Claim, that $V \cup W \in \mathrm{Max} (\mathcal{P}_{T} (A); \subset)$. 
Indeed, since $Q^{TT} = Q$, we have, in particular, $Q \subset (Q^{T})^{T}$, 
and so $Q \oplus Q^{T}$ is defined. Next, 
$Q \oplus Q^{T} = (V^{TT} \cup W^{TT})^{TT} = (V^{TTT} \cap W^{TTT})^{T} = (V^{T} \cap W^{T})^{T} = 
(V \cup W)^{TT}$. If $V \cup W$ is not maximal, then there exists $l_0 \in (V \cup W)^{T}$. 
At the same time, $(V \cup W)^{T} = (V \cup W)^{TTT} = (Q \oplus Q^{T})^{T} = (Q \cup Q^{T})^{TTT} = 
(Q \cap Q^{T})^{TT}$. But $Q \cap Q^{T} = \emptyset$, due to the first condition on $T$. 
Therefore, we continue: $(Q \cap Q^{T})^{TT} = \emptyset^{TT} = \emptyset$. 
Hence, $l_0$ cannot exist, and $V \cup W$ is maximal. 
Note, that we also have $(Q \oplus Q^{T})^{T} = \emptyset$, 
and as a corollary 
$Q \oplus Q^{T} = (Q \oplus Q^{T})^{TT} = \emptyset^{T} = A$.  

Now for the fourth axiom, take any $Q, Q_1, Q_2 \in \mathcal{P}^{T} (A)$, and assume that 
$Q \oplus Q_{1} = Q \oplus Q_{2}$. It is necessary to show, that $Q_1 = Q_2$. 
Represent $Q$, $Q_1$, and $Q_2$ in the form $Q = V^{TT}$, $Q_1 = V_{1}^{TT}$, and $Q_{2} = V_{2}^{TT}$, 
where $V, V_1, V_2 \in \mathcal{P}_{T} (A)$. Denote $Q_{0} := Q \oplus Q_{1} = Q \oplus Q_{2}$, and 
write it in the form $Q_{0} = U_{0}^{T}$, where $U_0 \in \mathcal{P}_{T} (A)$. 
Hence, $Q_{0}^{T} = U_{0}^{TT}$. We claim that both $(V \cup V_1) \cup U_0$ and $(V \cup V_2) \cup U_0$ 
are in $\mathrm{Max} (\mathcal{P}_{T} (A); \subset)$. 
Since $V \subset V^{TT} = Q \subset Q_{1}^{T} = V_{1}^{TTT} = V_{1}^{T}$, due to the first condition, 
the sets $V$ and $V_1$ are disjoint. Similarly, $V \cap V_2 = \emptyset$. 
We also have 
$Q_0 = Q \oplus Q_1 = (V^{TT} \cup V_{1}^{TT})^{TT} = (V^{TTT} \cap V_{1}^{TTT})^{T} = 
(V^{T} \cap V_{1}^{T})^{T} = (V \sqcup V_{1})^{TT}$, and $Q_{0}^{T} = U_{0}^{TT}$. 
Since $Q_{0} \oplus Q_{0}^{T}$ is defined, we similarly conclude that 
$V \sqcup V_1$ and $U_0$ are disjoint. Moreover, we already know, that in this case 
$(V \sqcup V_1) \sqcup U_0$ is maximal. 
Similarly, $(V \sqcup V_2) \sqcup U_0$ is maximal. 
Applying the main condition, one obtains: 
$V_{1}^{TT} = (V \cup U_{0})^{T} = V_{2}^{TT}$, i.e. $Q_{1} = Q_{2}$. 
Hence the fourth axiom is established. 

Consider the fifth axiom.
The candidate for $\mathbf{1}$ is $A$. 
It is easy to guess, that for $Q \in \mathcal{P}^{T} (A)$ it is necessary to put $Q^{*} := Q^{T}$. 
We already know, that $Q^{T} \oplus Q = A$, and since $A$ plays the role of $\mathbf{1}$, we obtain 
$Q^{*} \oplus Q = \mathbf{1}$. The fifth axiom is established. 

Finally, it remains to consider the sixth axiom. 
Note, that since $\mathbf{1} = A$, $\mathbf{0} = \emptyset$, and $A$ is not empty, 
one has $\mathbf{1} \not = \mathbf{0}$. 
Take any $Q \in \mathcal{P}^{T} (A)$, and assume that $Q \oplus \mathbf{1}$ is defined. 
This implies, that $Q \subset \mathbf{1}^{T} = A^{T} = \emptyset$. 
Hence, $Q = \emptyset$, i.e. $Q = \mathbf{0}$. 
The last axiom is established, and we have an effect algebra. 

It is not difficult to verify, that in fact this effect algebra is an orthoalgebra, 
and, moreover, a coherent orthoalgebra. 
Indeed, if we take any $Q \in \mathcal{P}^{T} (A)$, and assume, that 
$Q \oplus Q$ is defined, then this implies $Q \subset Q^{T}$. Hence, 
$Q = Q \cap Q^{T}$. But $Q \cap Q^{T} = \emptyset$ due to the first condition. 
Therefore $Q = \mathbf{0}$ ($\mathbf{0} := \emptyset$), i.e. our effect algebra is an orthoalgebra. 
Now, consider $Q, Q_1, Q_2 \in \mathcal{P}^{T} (A)$, and assume, that 
$Q \oplus Q_1$, $Q_1 \oplus Q_2$, and $Q_2 \oplus Q$ are defined. 
We have $Q \subset Q_{1}^{T}$, and $Q \subset Q_{2}^{T}$. 
Hence, $Q \subset Q_{1}^{T} \cap Q_{2}^{T}$. 
Apply the $(\cdot)^{T}$ operation: 
$Q^{T} \supset (Q_{1}^{T} \cap Q_{2}^{T})^{T} = (Q_{1} \cup Q_{2})^{TT} = Q_{1} \oplus Q_{2}$. 
Therefore $Q \oplus (Q_{1} \oplus Q_{2})$ is defined. The orthoalgebra is coherent. 
\qed

\section{The group of symmetry}

We have just three conditions on $T \subset A \times A$, which, when valid, 
allow to construct a coherent orthoalgebra. 
The first two are very simple, but the 
verification of the third one (the main condition), may be non-trivial. 
The main problem is, that there can be many elements in 
$\mathrm{Max} (\mathcal{P}_{T} (A); \subset)$. First, it is necessary to 
\emph{characterize} them all, and then, for every $M \in \mathrm{Max} (\mathcal{P}_{T} (A); \subset)$ 
and every $B \subset M$ verify the property $B^{T} = (M \backslash B)^{TT}$. 
A straightforward computation can become very complicated. 

The general approach to deal with this problem is to 
find some group of symmetry of $A$. 
Look at all bijections $\beta : A \os{\sim}{\to} A$, which respect the 
$T$ relation on $A$, i.e. $\forall l, l_1 \in A : (l_1, l) \in T \Rightarrow 
(\beta (l_1), \beta (l)) \in T$. 
Denote the group of all such bijections as $\mathit{Bij}_{T} (A)$. 
Every $\beta \in \mathit{Bij}_{T} (A)$ induces a bijective map 
from $\mathrm{Max} (\mathcal{P}_{T} (A); \subset)$ to itself. 
Suppose, we are able to describe some subgroup $\mathcal{G} \subset \mathit{Bij}_{T} (A)$, 
such that its natural action on $\mathrm{Max} (\mathcal{P}_{T} (A); \subset)$ has ``large'' orbits. 
Since it suffices to pick from each orbit just one representative, and verify 
the main condition on $T$ only for these, 
the verification of the main condition becomes more feasible. 

Let us now describe $A$, $T$, and $\mathcal{G}$ for the examples given below. 
Note, that these constructions clarify the combinatorics of the formulae 
present in \cite{RuugeFVO}. 
Let $V$ be a finite set, such that $N := \# V$ is divisible by $4$. 
Our construction will involve two collections of parameters with 
values in $\mathbb{Z}/ 2$. 
The first collection is 
indexed by $U \in \mathcal{P} (V)$ and the corresponding parameters 
are denoted as $b_{U} \in \mathbb{Z} / 2$. 
The second collection is indexed by 
$U, U_1 \in \mathcal{P} (V)$, $U \not = U_1$, 
and the parameters are denoted by $c_{U, U_1}$. 
It is assumed that $c_{U, U_1} = c_{U_1, U}$.  
Look at all maps $V \to \mathbb{Z} / 2$, and 
for every $U \in \mathcal{P} (V)$ denote 
\begin{equation} 
L_{b} (U) := \big\lbrace 
\phi : V \to \mathbb{Z} / 2 \, \big\vert \,
\sum_{v \in V} \phi (v) = b_{U}
\big\rbrace, 
\label{LbU_def} 
\end{equation}
where the index $b$ in the notation $L_{b} (\cdot)$ stands for $b := \lbrace b_{U} \rbrace_{U}$. 
Put 
\begin{equation} 
A_{b} := \bigsqcup_{U \in \mathcal{P} (V)} L_{b} (U). 
\label{Ab_def}
\end{equation}
Denote by $i_{U}^{b} : L_b (U) \rightarrowtail A_b$, $U \in \mathcal{P} (V)$, 
the canonical injections. 
Now define some relation $T_c$ on $A_b$, making use of the second 
collection of parameters $c := \lbrace c_{U, U_1} \rbrace_{U, U_1}$. 
For any $U, U_1 \in \mathcal{P} (V)$, $U \not = U_1$, and 
any $\phi, \phi' \in L (U)$ and $\phi_1 \in L (U_1)$, put
\begin{equation}
\begin{gathered} 
\big( i_U^b (\phi), i_U^b(\phi') \big) \in T_c \quad :\Leftrightarrow \quad \phi \not = \phi', \\
(\phi, \phi_1) \in T_c \quad :\Leftrightarrow \quad 
\sum_{s \in U \Delta U_1} 
\big( \phi (z) + \phi_1 (z) \big) \not = c_{U, U_1}, 
\end{gathered}
\label{Tc_def}
\end{equation}
where $\Delta$ denotes the symmetric difference of two subsets. 

We are going to apply with respect to $(A_b, T_c)$
the general construction of the previous section, i.e. 
substitute $A = A_{b}$, $T = T_{c}$, and try to adjust the parameters $b_{U}$ and $c_{U, U_1}$ 
in order to satisfy the three conditions.  
The main result of the present paper can now be outlined as follows: 
if the number of points $N$ in $V$ is divisible by $4$, then 
it is possible to choose the parameters $b_U$ and $c_{U, U_1}$ in such 
a way, that the assumptions of the proposition above are satisfied. 
Hence a new family of orthoalgebras is constructed. 
Moreover, it is possible to choose $c_{U, U_1}$ and $b_{U}$ in such a way, 
that the corresponding orthoalgebras do not admit bivaluations. 
It is interesting to stress the observed periodicity by $4$. 
Without $4 \vert N$, the construction does not work. 

Let us describe the group $\mathcal{G}$. 
Note, that the set of functions from $V$ to $\mathbb{Z} / 2$ 
may be viewed as a $N$-dimensional vector space over a field with two elements $\mathbb{F}_2$. 
Denote this vector space by $\mathbb{F}_{2}^{N}$. 
The number of elements in $\mathbb{F}_{2}^{N}$ is $2^{N}$. 
The sum of two vectors corresponds to a symmetric difference of two subsets. 
Look at the group of all automorphisms of this vector space, i.e. 
the general linear group $GL (N, \mathbb{F}_2)$ of $N \times N$ matrices with 
coefficients in $\mathbb{F}_2$. Let us describe a system of generators of this group 
(not a minimal one). For every $S \in \mathcal{P} (V)$, define a map 
$T_S : \mathcal{P} (V) \to \mathcal{P} (V)$, 
\begin{equation} 
 T_S (U) := 
\begin{cases} 
U & \text{if $\# (U \cap S)$ is even}, \\
\overline{U \Delta S} & \text{if $\# (U \cap S)$ is odd}, 
\end{cases}
\label{TS_def}
\end{equation}
where $U$ varies over $\mathcal{P} (V)$. 
Note, that these maps in case $N = 4$ have been introduced in \cite{RuugeFVO}. 
Hence, in order to compute $T_S (U)$, one needs to look at $S \cap U$. 
Observe, that $S \cap T_S (U) = S \cap U$. This implies, that $T_{S}^{2} = \textit{id}$. 
In particular, $T_S$ is a bijection. Moreover, 
for all $S, U, U_1 \in \mathcal{P} (V)$, we have 
\begin{equation*}
T_S (U \Delta U_1) = T_S (U) \Delta T_S (U_1). 
\end{equation*}
In order to prove the latter formula, note, that 
\begin{equation*}
\# \big( S \cap (U \Delta U_1) \big) = \# (S \cap U) + \# (S \cap U_1) - 2 \# (S \cap U \cap U_1). 
\end{equation*}
Therefore $\#_2 (S \cap (U \Delta U_1))$ ($\#_2$ denotes the cardinality of a set viewed in $\mathbb{Z}/ 2$) 
is determined by $\#_2 (S \cap U)$ and $\#_2 (S \cap U_1)$. 
Hence the $T_S$ correspond to \emph{linear} bijective maps of $\mathbb{F}_{2}^{N}$, 
i.e. $T_S$ corresponds to an element $\wh{T}_{S} \in GL (N, \mathbb{F}_2)$. 
The range of possible values of $S$ -- the set $\mathcal{P} (V)$ -- may be identified with $\mathbb{F}_{2}^{N}$.  
We denote by $\vert S \rangle$ an element of $\mathbb{F}_{2}$ corresponding to $S$. Note, that there exists a formula 
$\wh{T}_S \vert U \rangle = \vert T_S (U) \rangle$, 
where $U, S \in \mathcal{P} (V)$.

\begin{prop}
Let $T_{S}$, $\vert S \rangle \in \mathbb{F}_{2}^{N}$, be the set of  
reflections defined by the formula \eqref{TS_def}. 
Then $\lbrace \wh{T}_S \rbrace_S$ generates the whole group $GL (N, \mathbb{F}_2)$. 
\end{prop} 
\emph{Proof}. 
For particular $N$ small enough it is easy to verify the statement on 
computer in GAP. Let us provide a proof for all $N$. 
Note, that $T_{\emptyset} = T_{V} = \textit{id}$. 
Take any $S \in \mathcal{P} (V)$, $S \not = \emptyset, V$, 
and select $w \in \ov{S} := V \backslash S$. 
There is a useful formula: 
\begin{equation} 
\big(T_S \, T_{S \cup \lbrace w \rbrace} \, T_S \big) (\lbrace v \rbrace) = 
\begin{cases} 
\lbrace v \rbrace, & \text{if $v \not = w$}, \\
\ov{S}, & \text{if $v = w$}.  
\end{cases} 
\label{TSTSTS}
\end{equation}
It allows to prove (by induction) that the standard basis in $\mathbb{F}_{2}^{N}$ 
transforms into any other basis by a sequence of $\wh{T}_{S}$. 
Hence, the group is indeed $GL (N, \mathbb{F}_2)$. 
\qed

Note, that the fact that $\wh{T}_S$ are reflections, and the fact that they generate the whole  
general linear group, is obtained without using the assumption $4 \vert N$. 

We shall describe some bijections $A_{b} \os{\sim}{\to} A_{b}$, which respect the relation 
$T_c \subset A_b \times A_b$. The group $\mathcal{G}$ will be generated by these bijections. 
Before considering the general case, first look at the case where all the parameters 
$b_U$ and $c_{U, U_1}$ are put to $0 \in \mathbb{Z} / 2$. 
Write $A_0$ and $T_0$ in this case instead of $A_b$ and $T_c$, respectively. 
For every $S \in \mathcal{P} (V)$, define the maps 
$\theta_S : \mathrm{Maps} (V, \mathbb{Z} / 2) \to \mathrm{Maps} (V, \mathbb{Z} / 2)$ 
by the formulae: 
\begin{equation} 
\theta_S (\varphi) (v) := 
\begin{cases} 
\varphi (v), & \text{if $v \in \ov{S}$}, \\
\varphi (v) + \sum_{w \in \ov{S}} \varphi (w), & \text{if $v \in S$}, 
\end{cases}
\label{thetaS}
\end{equation} 
where $\varphi : V \to \mathbb{Z} / 2$, $\ov{S} := V \backslash S$. 
The latter can be expressed more compactly by  
$\theta_S (\varphi) (v) = \sum_{z \in T_S (\lbrace v \rbrace)} \varphi (z)$. 
A straightforward computation shows, that 
\begin{equation*} 
\theta_S^2 = \mathit{id}, 
\end{equation*}
and that for any $S, U \in \mathcal{P} (V)$, and any $\varphi: V \to \mathbb{Z} / 2$,  
the following formula is valid: 
\begin{equation} 
\sum_{v \in T_S (U)} \theta_S (\varphi) (v) = 
\sum_{v \in U} \varphi (v). 
\label{thetaSphiphi}
\end{equation}
This implies for any $U, S \in \mathcal{P} (V)$, that  
$\varphi \in L_0 (U)$ yields $\theta_S (\varphi) \in L_0 (T_S (U))$. 
It means that there exist induced maps 
\begin{equation*} 
\theta_{S, U} : L_0 (U) \to L_{0} (T_S (U)). 
\end{equation*}
For every $S \in \mathcal{P} (V)$, 
the collection $\lbrace \theta_{S, U} \rbrace_{U}$, $U \in \mathcal{P} (V)$, 
defines a bijective map 
\begin{equation*}
\wh{\theta}_{S} : A_{0} \os{\sim}{\to} A_{0}, 
\end{equation*}
(recall, $A_{0} = \sqcup_{U \in \mathcal{P} (V)} L_0 (U)$). 
The bijectivity follows from $\theta_S^2 = \mathit{id}$. 
Of course, $\wh{\theta}_{S}^{2} = \mathit{id}$ itself as well.  
Invoking that for any $S, U, U_1 \in \mathcal{P} (V)$, we have 
$T_S (U \Delta U_1) = T_S (U) \Delta T_S (U_1)$, 
it is not difficult to verify that all $\wh{\theta}_S$ respect the relation 
$T_0 \subset A_{0} \times A_{0}$, or, equivalently, the 
relation $(A_0 \times A_0) \backslash T_0$. 
Consider $U, U_1 \in \mathcal{P} (V)$ and $\varphi \in L_0 (U)$, 
$\varphi_1 \in L_0 (U_1)$, such that $(i_{U}^{0} (\varphi), i_{U_1}^{0} (\varphi_1)) \not \in T_{0}$ 
($i_{U}^{0}$ and $i_{U_1}^{0}$ denote the canonical injections into $A_0$). 
If $U_1 = U$, then the fact mentioned is implied by the bijectivity of $\wh{\theta}_S$. 
If $U \not = U_1$, then we have 
$\sum_{v \in U \Delta U_1} \big( \varphi (v) + \varphi_1 (v) \big) = 0$. 
Therefore,  
$\sum_{v \in T_S (U \Delta U_1)} \big( \theta_S (\varphi) (v) + 
\theta_S (\varphi_1) (v) \big) = 0$. 
Since $T_S (U \Delta U_1) = T_S (U) \Delta T_S(U_1)$, the pair of elements in $A_0$ 
that correspond to $\theta_S (\varphi) \in L_0 (T_S (U))$ and 
$\theta_S (\varphi_1) \in L_0 (T_S (U_1))$, is in relation $(A_0 \times A_0) \backslash T_0$.  

Now let us generalize the construction of the maps $\wh{\theta}_S$. 
We have the collections of parameters $b = \lbrace b_U \rbrace_{U}$, and 
$c = \lbrace c_{U, U_1} \rbrace_{U, U_1}$. 
For every $U, S \in \mathcal{P} (V)$, we need to describe some maps 
$L_b (U) \to L_b (T_S (U))$. In the case considered above, these were the maps $\theta_{S, U}$. 
For every fixed $S$, the whole collection $\lbrace \theta_{S, U} \rbrace_{U}$ stemmed 
just from one ``global'' function $\theta_S$. 
Now, let us not assume this property. Take an arbitrary collection of $\mathbb{Z} / 2$-valued 
parameters $\lbrace a_{S, U} (v) \rbrace_{S, U, v}$, $S, U \in \mathcal{P} (V)$, $v \in V$, 
and try to define some maps 
$\theta_{S, U}^{(a)} : L_b (U) \to L_b (T_S (U))$ by the formula 
\begin{equation} 
\theta_{S, U}^{(a)} (\varphi) (v) := \theta_{S} (\varphi) (v) + a_{S, U} (v),  
\label{thetaSUa}
\end{equation} 
where $\varphi \in L_b (U)$, $v \in V$. 
The case considered above corresponds to all $a_{S, U} (v) = 0$. 
It is necessary to ensure, that $\theta_{S, U}^{(a)} (\varphi) (v) \in L_b (T_S (U))$. 
This yields a condition on $a_{S, U} (v)$: 
\begin{equation*}
\sum_{v \in U} \theta_{S, U}^{(a)} (\varphi) (v) = b_{T_S (U)},  
\end{equation*}
where $\varphi \in L_b (U)$. 
Expanding the definitions of $L_b (U)$ and $\theta_{S, U}^{(a)}$, one reduces 
this equality just to  
$0 = b_{T_S (U)}$, if $\# (S \cap U)$ is even, and to 
$b_{U} + \sum_{v \in \ov{U \Delta S}} a_{S, U} (v) = b_{T_S (U)}$, if $\# (S \cap U)$ is odd. 
Both cases are captured by one formula: 
\begin{equation} 
\sum_{v \in T_S (U)} a_{S, U} (v) = b_{T_S (U)} + b_{U}, 
\label{aSU_bb}
\end{equation}
where $S$ and $U$ vary over $\mathcal{P} (V)$. Assume, that this condition is satisfied. 
Hence, we have well-defined maps $\theta_{S, U}^{(a)} : L_b (U) \to L_b (T_S (U))$. 
Since $\theta_{S, U}$ are bijections, so are $\theta_{S, U}^{(a)}$. 
For every fixed $S \in \mathcal{P} (V)$, the collection 
$\lbrace \theta_{S, U}^{(a)} \rbrace_{U}$ defines a bijective map 
\begin{equation*}
 \wh{\theta}_{S}^{(a)} : A_{b} \os{\sim}{\to} A_{b}. 
\end{equation*}
Impose a requirement, that $\wh{\theta}_{S}^{(a)}$ respects the relation $T_c \subset A_b \times A_b$. 
This yields another condition on the parameters $a_{S, U} (v)$. 
Take any $U, U_1 \in \mathcal{P} (V)$. Since $\wh{\theta}_{S}^{(a)}$ is bijective, 
the requirement is satisfied if $U_1 = U$. Let $U_1 \not = U$. 
Take any $\varphi \in L_b (U)$, $\varphi_1 \in L_b (U_1)$, and assume that 
$\sum_{v \in U \Delta U_1} (\varphi (v) + \varphi_1 (v)) = c_{U, U_1}$. 
This should imply 
$\sum_{v \in T_S (U) \Delta T_S (U_1)} 
( \theta_{S, U}^{(a)} (\varphi) (v) + 
\theta_{S, U_1}^{(a)} (\varphi_1) (v) ) = c_{T_S (U), T_S (U_1)}$. 
Taking into account, that $T_S (U) \Delta T_S (U_1) = T_S (U \Delta U_1)$, 
expanding the definitions \eqref{thetaSUa} of $\theta_{S, U}^{(a)}$ and $\theta_{S, U_1}^{(a)}$, 
and taking into account the mentioned formula \eqref{thetaSphiphi} for $\theta_S$, 
one reduces this requirement to the form 
\begin{equation} 
\sum_{v \in T_S (U \Delta U_1)} 
\big( 
a_{S, U} (v) + a_{S, U_1} (v) 
\big) = c_{U, U_1} + c_{T_S (U), T_S (U_1)}, 
\label{aSU_cc}
\end{equation}
where $S$, $U$, and $U_1$ vary over $\mathcal{P} (V)$, and $U_1 \not = U$. 

We have an overdetermined system of linear equations 
\eqref{aSU_bb}, \eqref{aSU_cc}, 
with respect to the 
indeterminates $a_{S, U} (v) \in \mathbb{Z} /2$. 
The quantities $b_U$ and $c_{U, U_1}$ are parameters. 
It is necessary to solve this system of equations, and then obtain a condition 
of solvability in terms of $b_U$ and $c_{U, U_1}$. After that 
$b_U$ and $c_{U, U_1}$ become indeterminates themselves, and 
one needs to find at least some solutions of the solvability equations. 
Assume all this is accomplished. Then we obtain a collection of bijective 
maps $\wh{\theta}_{S}^{(a)} : A_{b} \os{\sim}{\to} A_{b}$, which respect 
the relation $T_c$. They generate some group $\mathcal{G}_{a} \subset \mathit{Bij}_{T_c} (A_b)$. 
In what follows, it is this group that will be used to establish the main condition on $T_c$, that 
allows to construct the orthoalgebra. Moreover, that parameters $b_U$ and $c_{U, U_1}$ can be chosen 
in such a way, that the corresponding orthoalgebra does not admit a bivaluation (this is the easy part). 
 
\section{The solutions}

Let us rewrite the equation \eqref{aSU_cc} as follows. 
This equation contains a sum over $v \in T_S (U \Delta U_1)$. 
This is the same as the sum over $v \in T_S (U) \Delta T_S (U_1)$. 
Since the terms in this sum are $\mathbb{Z} / 2$-valued, 
it can be split as $\sum_{v \in T_S (U)} + \sum_{v \in T_S (U_1)}$.   
Perform this action upon the equation \eqref{aSU_cc}, and then 
use twice the equations \eqref{aSU_bb} 
corresponding to $U_0 = U$ and $U_0 = U_1$. 
It is convenient to denote 
\begin{gather*} 
c_{U, U_1}^{(S)} := c_{U, U_1} + b_{U} + b_{U_1} + c_{T_S (U), T_S (U_1)} + b_{T_S (U)} + b_{T_S (U_1)}, \\
b_{U_0}^{(S)} := b_{U_0} + b_{T_S (U_0)}. 
\end{gather*}
The system of equations \eqref{aSU_bb}, \eqref{aSU_cc}, is equivalent to:  
\begin{gather} 
\sum_{v \in T_S (U_1)} a_{S, U} (v) + 
\sum_{v \in T_S (U)} a_{S, U_1} (v) = 
c_{U, U_1}^{(S)}, 
\label{aa_cuus}
\\ 
\sum_{v \in T_S (U_0)} a_{S, U_0} (v) = b_{U_0}^{(S)}. 
\label{aa_bus}
\end{gather}

Let us express all $a_{S, Q} (v)$ with $\# Q \geqslant 2$ via the indeterminates of the 
form $a_{S, \lbrace z \rbrace} (v)$. 
Let $Q \in \mathcal{P} (V)$ be any subset such that $\# Q \geqslant 2$, 
and $u \in S$ and $w \in \ov{S}$ be any points. 
Look at the equation \eqref{aa_cuus}. 
Put $U = Q$ and $U_1 = \pnt{w}$. This allows to find $a_{S, Q} (w)$: 
\begin{equation*} 
a_{S, Q} (w) = \sum_{v \in T_S (Q)} a_{S, \pnt{w}} (v) + 
c_{Q, \pnt{w}}^{(S)}, \quad w \in \ov{S}.  
\end{equation*}
Next, put $U = Q$ and $U_1 = \pnt{u}$. 
Since $T_S (\pnt{u}) = \pnt{u} \cup \ov{S}$, 
the resulting expression on the left-hand side will contain a sum of 
$a_{S, Q} (v)$ over $v \in \pnt{u} \cup \ov{S}$. 
For all values of $v$, except $v = u$, 
we already can express $a_{S, Q} (v)$. 
Hence, it is possible to find $a_{S, Q} (u)$: 
\begin{equation*}
a_{S, Q} (u) = 
\sum_{w' \in \ov{S}} 
a_{S, Q} (w') + 
\sum_{v \in T_S (Q \Delta \lbrace u \rbrace)} 
a_{S, \lbrace u \rbrace} (v) + 
c_{Q, \pnt{u}}^{(S)}, \quad u \in S. 
\end{equation*}

Now consider the case where the sets $U_0$, $U$, and $U_1$, are singletons. 
Let $u, u_1 \in S$ and $w, w_1 \in \ov{S}$ be any points. 
The equations \eqref{aa_bus} corresponding to $U_0 = \pnt{w}$ and $U_0 = \pnt{u}$, respectively, yield: 
\begin{gather*} 
a_{S, \pnt{w}} (w) = 0, \\
a_{S, \pnt{u}} (u) = b_{\pnt{u}}^{(S)} + 
\sum_{w' \in \ov{S}} a_{S, \pnt{u}} (w'). 
\end{gather*}
For the $c$-equations, it is necessary to consider the 
following three cases: 
1) $U = \pnt{w}$, $U_1 = \pnt{w_1}$; 
2) $U = \pnt{u}$, $U_1 = \pnt{w}$; 
3) $U = \pnt{u}$, $U_1 = \pnt{u_1}$.  
They yield: 
\begin{gather*} 
a_{S, \pnt{w}} (w_1) + a_{S, \pnt{w_1}} (w) = 0, \\
a_{S, \pnt{u}} (w) + a_{S, \pnt{w}} (u) = 
c_{\pnt{u}, \pnt{u_1}}^{(S)} + \sum_{w' \in \ov{S}} a_{S, \pnt{w}} (w'), \\
a_{S, \pnt{u}} (u_1) + a_{S, \pnt{u_1}} (u) = 
c_{\pnt{u}, \pnt{u_1}}^{(S)} + 
\sum_{w' \in \ov{S}} 
\Big( a_{S, \pnt{u}} (w') + a_{S, \pnt{u_1}} (w') \Big). 
\end{gather*} 
For every fixed $S \in \mathcal{P} (V)$, one may view the latter five equalities 
as a system of linear equations with respect to $a_{S, \pnt{z}} (v) \in \mathbb{Z} / 2$, $v, z \in V$.  
It is not difficult to verify, that the corresponding \emph{homogeneous} system of equations has 
many solutions. 
Redenote the indeterminates in this system as $\alpha_{S, \pnt{z}} (v)$, $v, z \in V$. 
Denote $\mathcal{E} (V) := \lbrace U \subset V \, | \, \# U = 2 \rbrace$. 
Write $vz$ instead of $\lbrace v, z \rbrace$ for the elements of $\mathcal{E} (V)$.  
Take any function $\mu : \mathcal{E} (V) \to \mathbb{Z} / 2$, and denote 
\begin{equation*} 
\chi_{\mu}^{Q} (v) := \sum_{z \in Q \backslash \pnt{v}} \mu (vz), 
\end{equation*}
where $Q \in \mathcal{P} (V)$, $v \in V$. 
It is not difficult to verify in a straightforward manner, that 
$\alpha_{\pnt{z}} (v) = \chi_{\mu}^{T_S (\pnt{z})} (v)$ 
defines a solution of the homogeneous system. 
We just remark, that 
$\chi_{\mu}^{T_S (\pnt{w})} (v) = \mu (vw)$ for $w \in \ov{S}$, 
$\chi_{\mu}^{T_S (\pnt{u})} (v) = \mu (uv) + \sum_{w' \in \ov{S}} \mu (vw')$ 
for $u \in S$, and it is convenient to accept a formal agreement $\mu (vv) = 0$ in order to 
perform this computation. 

We need a solution of the non-homogeneous system. 
Let $u, u_1 \in S$ and $w, w_1 \in \ov{S}$ be any points. 
Put 
\begin{equation*} 
\ov{a}_{S, \pnt{w}} (w_1) := 
\begin{cases} 
0, & \text{if $w \prec w_1$}, \\
c_{\pnt{w}, \pnt{w_1}}^{(S)}, & \text{if $w_1 \prec w$}, \\
b_{\pnt{w}}^{(S)}, & \text{if $w_1 = w$}; 
\end{cases}
\end{equation*}
\begin{equation*} 
\ov{a}_{S, \pnt{u}} (u_1) := 
\begin{cases} 
0, & \text{if $u \prec u_1$}, \\ 
c_{\pnt{u}, \pnt{u_1}}^{(S)}, & \text{if $u_1 \prec u$}, \\
b_{\pnt{u}}^{(S)}, & \text{if $u_1 = u$}; 
\end{cases}
\end{equation*}
\begin{gather*} 
\ov{a}_{S, \pnt{w}} (u) := 
b_{\pnt{w}}^{(S)} + c_{\pnt{u}, \pnt{w}}^{(S)} + 
\sum_{\substack{w' \in \ov{S}, \\ w' \prec w}} c_{\pnt{w'}, \pnt{w}}^{(S)}; \\
\ov{a}_{s, \pnt{u}} (w) := 0. 
\end{gather*}
A straightforward computation shows that $a_{S, \pnt{z}} (v) = \ov{a}_{S, \pnt{z}} (v)$ 
is a solution. 
Moreover, any other solution $a_{S, \pnt{z}} (v) = \wh{a}_{S, \pnt{z}} (v)$ 
can be represented in the form: 
\begin{equation*}
\wh{a}_{S, \pnt{z}} (v) = \ov{a}_{S, \pnt{z}} (v) + \chi_{\wh{\mu}}^{T_S (\pnt{z})} (v),
\end{equation*} 
for some $\wh{\mu} : \mathcal{E} (V) \to \mathbb{Z} /2$. 
For any $u, u \in S$, $u \not = u_1$, and 
$w, w_1 \in \ov{S}$, $w \not = w_1$, 
the values of $\wh{\mu} (ww_1)$, $\wh{\mu} (uu_1)$, and $\wh{\mu} (uw)$ are 
given by the formulae 
\begin{equation*} 
\wh{\mu} (ww_1) := 
\begin{cases} 
\wh{a}_{S, \pnt{w}} (w_1), & \text{if $w \prec w_1$}, \\
\wh{a}_{S, \pnt{w_1}} (w), & \text{if $w_1 \prec w$}; 
\end{cases}
\end{equation*} 
\begin{equation*} 
\wh{\mu} (uu_1) := 
\begin{cases} 
\wh{a}_{S, \pnt{u}} (u_1), & \text{if $u \prec u_1$}, \\
\wh{a}_{S, \pnt{u_1}} (u) + 
\sum_{w' \in \ov{S}} \wh{a}_{S, \pnt{u}} (w'), & \text{if $u_1 \prec u$}; 
\end{cases}
\end{equation*}
and 
\begin{equation*} 
\wh{\mu} (uw) := 
\wh{a}_{S, \pnt{u}} (w) + 
\sum_{\substack{w' \in \ov{S}, \\ w' \prec w}} 
\wh{a}_{S, \pnt{w'}} (w) + 
\sum_{\substack{w' \in \ov{S}, \\ w' \succ w}} 
\wh{a}_{S, \pnt{w}} (w'). 
\end{equation*}
The verification is straightforward. 
Therefore, \emph{any} solution of the homogeneous system is of the form 
$\alpha_{S, \pnt{z}} (v) = \chi_{\mu}^{T_S (\pnt{z})} (v)$, 
$\mu$ -- some function.    
One can now take a solution for $a_{S, \pnt{z}} (v)$, and compute 
the rest of of the $a_{S, Q} (v)$ according to the formulae derived above. 
Note, that the transformation 
$\alpha_{\pnt{z}} (v) = \chi_{\mu}^{T_S (\pnt{z})} (v) \to 
\alpha_{\pnt{z}} (v) = \chi_{\mu}^{T_S (\pnt{z})} (v) + 
\chi_{\mu}^{T_S (\pnt{z})} (v)$ induces the transformation of $a_{S, Q} (v)$ of the form: 
$a_{S, Q} (v) \to a_{S, Q} (v) + \chi_{\mu}^{T_S (Q)} (v)$, 
i.e. there is a gauge symmetry group of transformations 
for the system of equations for $a_{S, Q} (v)$.  

We have the expressions for all $a_{S, Q} (v)$, but we did not use all the equations of the system. 
Take any $S \in \mathcal{P} (V)$, and any 
$Q, Q_1 \in \mathcal{P} (V)$, $Q_1 \not = Q$. 
Substituting these expressions into the equations, one obtains the conditions: 
\begin{multline*} 
c_{Q, Q_1}^{(S)} + \sum_{z_1 \in Q_1} c_{Q, \pnt{z_1}}^{(S)} + 
\sum_{z \in Q} c_{\pnt{z}, Q_1}^{(S)} = \\ =
\sum_{z_1 \in Q_1} \sum_{v \in T_S (Q)} 
a_{S, \pnt{z_1}} (v) + 
\sum_{z \in Q} \sum_{v_1 \in T_S (Q_1)} 
a_{S, \pnt{z}} (v_1), 
\end{multline*}  
and 
\begin{equation*} 
b_{Q}^{(S)} + \sum_{z \in Q} c_{Q, \pnt{z}}^{(S)} = 
\sum_{z \in Q} \sum_{v \in T_S (Q)} a_{S, \pnt{z}} (v).  
\end{equation*}
Denote the right-hand sides of these equalities by 
$X_a (S, Q, Q_1)$ and $Y_a (S, Q, Q_1)$, respectively. 
Note, that these two quantities are invariant under the gauge 
transformation $a_{S, \pnt{z}} (v) \to a_{S, \pnt{z}} (v) + 
\chi_{\mu}^{T_S (\pnt{z})} (v)$, ($\mu$ -- any function). 
It remains to substitute $a_{S, \pnt{z}} (v) = \ov{a}_{S, \pnt{z}} (v)$ and compute 
the corresponding $X_{\ov{a}}$ and $Y_{\ov{a}}$. 

In order to compute $Y_{\ov{a}}$ it is necessary to consider two cases: 
$\# (Q \cap S)$ is even, and $\# (Q \cap S)$ is odd. 
The computation in the first case is a little bit easier, but it turns out, that 
in both cases the result is the same: 
\begin{equation*} 
Y_{\ov{a}} (S, Q, Q_1) = 
\sum_{z \in Q} b_{\pnt{z}}^{(S)} + 
\sum_{\substack{z, z' \in Q, \\ z \prec z'}} c_{\pnt{z}, \pnt{z'}}^{(S)}. 
\end{equation*}
The value of the sum on the right-hand side does not depend on $\prec$, 
due to the symmetry $c_{U, U_1}^{(S)} = c_{U_1, U}^{(S)}$, which is implied 
by the assumption $c_{U, U_1} = c_{U_1, U}$. 

In order to compute $X_{\ov{a}} (S, Q, Q_1)$, it is necessary to investigate the 
following three cases: 
1) both $\# (Q \cap S)$ and $\# (Q_1 \cap S)$ are even; 
2) $\# (Q \cap S)$ is odd, and $\# (Q_1 \cap S)$ is even; 
3) both $\# (Q \cap S)$ and $\# (Q_1 \cap S)$ are odd. 
In all three cases, one obtains the same expression: 
\begin{equation*}
X_{\ov{a}} (S, Q, Q_1) = 
\sum_{\substack{
z \in Q, \, z_1 \in Q_1, \\ z \not = z_1
}} c_{\pnt{z}, \pnt{z_1}}^{(S)}. 
\end{equation*}
Therefore, we obtain the following conditions: 
\begin{equation} 
c_{Q, Q_1}^{(S)} + \sum_{z_1 \in Q_1} c_{Q, \pnt{z_1}}^{(S)} + 
\sum_{z \in Q} c_{\pnt{z}, Q_1}^{(S)} + 
\sum_{\substack{
z \in Q, \, z_1 \in Q_1, \\ z \not = z_1
}} c_{\pnt{z}, \pnt{z_1}}^{(S)} = 0, 
\label{c_cond}
\end{equation} 
and 
\begin{equation} 
\sum_{z \in Q} c_{Q, \pnt{z}}^{(S)} + 
\sum_{\substack{
z, z' \in Q, \\ z \prec z' 
}} c_{\pnt{z'}, \pnt{z}}^{(S)} = 
b_{Q}^{(S)} + \sum_{z \in Q} b_{\pnt{z}}^{(S)}. 
\label{b_cond}
\end{equation}
Recall that $S$, $Q$, and $Q_1$ vary over $\mathcal{P} (V)$, 
$Q_1 \not = Q$. By definition, we put formally $c_{Q, Q}^{(S)} = 0$. 
Note, that if $\# Q = 1$, then the second condition \eqref{b_cond} turns into an identity. 
Similarly, if at least one of the sets $Q$ or $Q_1$ has cardinality $1$, then 
the first condition \eqref{c_cond} trivializes as well. 
These two conditions are the conditions of the solvability of the system of equations for 
$\lbrace a_{S, U} (v) \rbrace_{v, U, S}$. 

\section{Periodicity by four}

Is it possible to satisfy the obtained solvability conditions \eqref{b_cond}, \eqref{c_cond}? 
We shall not try to describe all the solutions, but construct some. 
The crucial assumption is the following. 
Let us search for $c_{U, U_1}$ and $b_{U}$ in the form 
\begin{equation}
\begin{gathered} 
c_{U, U_1} = c(\#_4 (U \Delta U_1)), \\
b_{U} = b (\#_4 U), 
\end{gathered}
\label{bc_four}
\end{equation}
where $U, U_1 \in \mathcal{P} (V)$, $U_1 \not = U$, 
$\#_4 (\cdot)$ denotes the cardinality of the subset viewed in $\mathbb{Z} / 4$, and 
$c (\cdot) : \mathbb{Z} / 4 \to \mathbb{Z} / 2$ and 
$b (\cdot) : \mathbb{Z} / 4 \to \mathbb{Z} / 2$ are unknown functions. 
A not quite trivial property of the solvability 
system of equations \eqref{c_cond}, \eqref{b_cond}, is 
that it admits such an anzats if the number of points $N$ in $V$ is divisible by $4$.   

Take any $S$, and look at the quantity $b_{\pnt{z}}^{(S)}$, $z \in V$. 
Observe, that since $b_U = b (\#_4 U)$, its value depends only on 
whether $z \in S$ or $z \not \in S$. In other words, one may take \emph{any} 
$u_0 \in S$ and $w_0 \in \ov{S}$, and claim that 
$b_{\pnt{z}}^{(S)} = b_{\pnt{u_0}}^{(S)}$, if $z \in S$, and 
$b_{\pnt{z}}^{(S)} = b_{\pnt{w_0}}^{(S)}$, if $z \in \ov{S}$.  
Similar statements may be made about the quantities of the form 
$c_{Q, \pnt{z_1}}^{(S)}$, $c_{\pnt{z}, Q_1}^{S}$, and $c_{\pnt{z}, \pnt{z_1}}^{(S)}$. 

Choose any $S$, and  $Q$, $Q_1$ such that $Q_1 \not = Q$. 
Look at the set $S$. 
It gets partitioned into four subsets: 
\begin{equation*} 
S = (S \cap Q \cap Q_1) \sqcup (S \cap \ov{Q} \cap Q_1) \sqcup 
(S \cap Q \cap \ov{Q}_1) \sqcup (S \cap \ov{Q} \cap \ov{Q}_1). 
\end{equation*}
In each of the subsets, if non-empty, 
choose a point (it doesn't matter which one): 
$\xi_0 \in S \cap Q \cap Q_1$, 
$\xi_1 \in S \cap \ov{Q} \cap Q_1$, 
$\xi_2 \in S \cap Q \cap \ov{Q}_1$, and 
$\xi_3 \in S \cap \ov{Q} \cap \ov{Q}_1$. 
Denote the cardinalities of these four subsets by 
$m_0$, $m_1$, $m_2$, and $m_3$, respectively. 
Next, perform a similar process with respect to $\ov{S}$, i.e. choose arbitrary four points 
$\eta_0$, $\eta_1$, $\eta_2$, and $\eta_3$, such that 
$\eta_0 \in \ov{S} \cap Q \cap Q_1$, 
$\eta_1 \in \ov{S} \cap \ov{Q} \cap Q_1$, 
$\eta_2 \in \ov{S} \cap Q \cap \ov{Q}_1$, and 
$\eta_3 \in \ov{S} \cap \ov{Q} \cap \ov{Q}_1$. 
(If a set is empty, the corresponding point will not be needed).  
Denote the cardinalities of these subsets as $n_0$, $n_1$, $n_2$, and $n_3$, respectively. 
Note, that $c_{\pnt{z}, \pnt{z_1}}^{(S)} = 0$, if both $z$ and $z_1$ are in $S$, or both are in $\ov{S}$. 
With this remark, the solvability equation \eqref{c_cond} after the described anzats, acquires the form: 
\begin{multline*} 
c_{Q, Q_1}^{(S)} + 
\big( 
m_0 \, c_{Q, \pnt{\xi_0}}^{(S)} + 
n_0 \, c_{Q, \pnt{\eta_0}}^{(S)} + 
m_1 \, c_{Q, \pnt{\xi_1}}^{(S)} + 
n_1 \, c_{Q, \pnt{\eta_1}}^{(S)} 
\big) + \\ + 
\big[ 
m_0 \, c_{Q_1, \pnt{\xi_0}}^{(S)} + 
n_0 \, c_{Q_1, \pnt{\eta_0}}^{(S)} + 
m_2 \, c_{Q_1, \pnt{\xi_2}}^{(S)} + 
n_2 \, c_{Q_1, \pnt{\eta_2}}^{(S)}
\big] + \\ + 
\big\lbrace 
m_0 \, n_1 + n_0 \, m_1 + 
m_0 \, n_2 + n_0 \, m_2 +  
m_1 \, n_2 + n_1 \, m_2  
\big\rbrace c_{\pnt{\xi_3}, \pnt{\eta_3}}^{(S)} = 0. 
\end{multline*}
Note, that the values of $m_i$ and $n_i$ ($i = 0, 1, 2, 3$) 
depend on the sets $S$, $Q$, and $Q_1$.  
Of course, $\sum_{i = 0}^{3} (m_i + n_i) = N$. 
Note, that it suffices to know 
only the images of $m_i$ and $n_i$ ($i = 0, 1, 2, 3$) in $\mathbb{Z} / 2$. 

The solvability condition \eqref{b_cond} is reduced in a similar way. 
This time we do not need the set $Q_1$. 
Take any $S$ and $Q$, 
choose any points 
$\zeta \in Q \cap S$, $\omega \in Q \cap \ov{S}$, 
and then any 
$\zeta' \in Q \cap S$, $\zeta' \not = \zeta$, and 
$\omega' \in Q \cap \ov{S}$, $\omega' \not = \omega$, 
(if some of these points cannot be chosen, they are not needed). 
Denote $k := \# (Q \cap S)$ and $l := \# (Q \cap \ov{S})$. 
The condition reduces to the form: 
\begin{multline*} 
k \, c_{Q, \pnt{\zeta}}^{(S)} + 
l \, c_{Q, \pnt{\omega}}^{(S)} + 
\frac{k (k - 1)}{2} c_{\pnt{\zeta}, \pnt{\zeta'}}^{(S)} + 
\frac{l (l - 1)}{2} c_{\pnt{\omega}, \pnt{\omega'}}^{(S)} + \\ + 
k \, l \, c_{\pnt{\zeta}, \pnt{\omega}}^{(S)} = 
b_{Q}^{(S)} + k \, b_{\pnt{\zeta}}^{(S)} + l \, b_{\pnt{\omega}}^{(S)}.  
\end{multline*}
Note, that each time, when the corresponding points cannot be chosen, 
the term that contains this point contains a factor equal to zero. 
The values of $k$ and $l$ depend on the sets $S$ and $Q$. 
Note that it suffices to know only the image of $l$ in $\mathbb{Z} / 2$, and 
the image of $k$ in $\mathbb{Z} / 4$ (not $\mathbb{Z} / 2$)! 

It remains to perform the mentioned anzats in these equations 
and simplify them. 
It is convenient to use the following formulae: 
\begin{gather*}
\#_4 (U \Delta U_1) = \#_4 U + \#_4 U_1 - 2 \#_4 (U \cap U_1), \\
\forall i \in \mathbb{Z}/ 4: [i]_2 = 0 \Rightarrow 2 i = 0, \\
\forall i \in \mathbb{Z}/ 4: [i]_2 = 1 \Rightarrow 2 i = 2, \\ 
\end{gather*}
where $[i]_2$ denotes the canonical image of $i$ in $\mathbb{Z}/ 2$, 
$U$ and $U_1$ are any subsets of $V$. 
We shall also need the assumption that the number $N$ of points in $V$ is divisible by $4$. 
In this case, for all $U \in \mathcal{P} (V)$, the following formula is valid: 
\begin{equation*} 
\#_4 \ov{U} = - \#_4 U. 
\end{equation*}  

First look at the equation \eqref{b_cond}. 
Recall, that $\zeta \in S \cap Q$, and $\omega \in \ov{S} \cap Q$. 
We have: 
\begin{gather*} 
b_{Q}^{(S)} = b (\#_4 Q) + b (\#_4 T_S (Q)), \\
b_{\pnt{\zeta}}^{(S)} = b (\#_4 \pnt{\zeta}) + b (\#_4 T_S (\pnt{\zeta})) = 
b (1) + b ( - \#_4 S + 1), \\
b_{\pnt{\omega}}^{(S)} = b (\#_4 \pnt{\omega}) + b (\#_4 T_S (\pnt{\omega})) = 0. 
\end{gather*}
Similar computations yield: 
\begin{gather*} 
c_{Q, \pnt{\zeta}}^{(S)} = 
b_{Q}^{(S)} + b_{\pnt{\zeta}}^{(S)} + 
c(\#_4 Q - 1) + c (\#_4 T_S (Q \Delta \pnt{\zeta})), \\
c_{Q, \pnt{\omega}}^{(S)} = 
b_{Q}^{(S)} + b_{\pnt{\omega}}^{(S)} +  
c (\#_4 Q - 1) + c (\#_4 T_S (Q \Delta \pnt{w})), \\
c_{\pnt{\zeta}, \pnt{\zeta'}}^{(S)} = 0, \\
c_{\pnt{\zeta}, \pnt{\omega}}^{(S)} = b (1) + b (- \#_4 S + 1) + c (2) + c (- \#_4 S), \\
c_{\pnt{\omega}, \pnt{\omega'}}^{(S)} = 0. 
\end{gather*}
We need to compute $\#_4 T_S (Q)$, $\#_4 T_S (Q \Delta \pnt{\zeta})$, and 
$\#_4 T_S (Q \Delta \pnt{w})$. 
Put:  
\begin{equation*} 
s := \#_4 S, \quad 
q := \#_4 Q, \quad 
t := \#_4 (Q \cap S).
\end{equation*}
If $\# (Q \cap S)$ is even, (i.e. $t = 0, 2$), then 
\begin{gather*} 
\#_4 T_S (Q) = q, \\
\#_4 T_S (Q \Delta \pnt{\zeta}) = - s - q - 1, \\ 
\#_4 T_S (Q \Delta \pnt{w}) = q - 1. 
\end{gather*}
If $\# (Q \cap S)$ is odd, (i.e. $t = 1, 3$), then 
\begin{gather*} 
\#_4 T_S (Q) = - s - q + 2, \\
\#_4 T_S (Q \Delta \pnt{\zeta}) = q - 1, \\ 
\#_4 T_S (Q \Delta \pnt{w}) = -s - q - 1. 
\end{gather*}
Hence, it suffices to know 
the values of thee parameters $s, q, t \in \mathbb{Z} / 4$ in order 
to compute the left and right-hand expressions of the equation \eqref{b_cond}. 
(Of course, $[k]_4 = t$, and $[l]_4 = q - t$.) 
It turns out (this can be easily verified on a computer 
in Maple, or by a straightforward computation), 
that for each of the $4^3$ possible variants of $(s, q, t)$, the reduced equation acquires only 
one of the following types: 
either it becomes an identity $0 = 0$, or one of the two equations 
\begin{equation}
\begin{gathered} 
c (0) + c (2) = b (0) + b (2), \\
c (1) + c (3) = b (1) + b (3), 
\end{gathered} 
\label{cc_bb}
\end{equation}
or their sum $\sum_{i = 0}^{3} (c (i) + b (i)) = 0$. 
One may assign arbitrary values, say to all $c(i)$ and to $b (0)$, $b (1)$, and then 
determine $b (2)$ and $b (3)$. 

The equations \eqref{c_cond} are reduced in a similar way, and in the final stage it  
is best to compute in Maple. Let us describe all the preparatory work. 
Look at $c_{Q, Q_1}^{(S)}$. 
We have $c_{Q, Q_1}^{(S)} = b_{Q}^{(S)} + b_{Q_1}^{(S)} + 
c (\#_4 (Q \Delta Q_1)) + c (\#_4 T_S (Q \Delta Q_1))$. 
In particular, it is necessary to know $\#_2 S \cap (Q \Delta Q_1)$. 
Since $\# S \cap (Q \Delta Q_1) = \# (S \cap Q) + \# (S \cap Q_1) - 2 \# (S \cap Q \cap Q_1)$, 
and the latter term is even, one has 
\begin{equation*}
\#_{2} S \cap (Q \Delta Q_1) = \#_{2} (S \cap Q) + \#_{2} (S \cap Q_1). 
\end{equation*}
Denote 
\begin{gather*}
s := \#_4 S, \quad q := \#_4 Q, \quad q_1 := \#_4 Q_1, \\
t := \#_4 (S \cap Q), \quad t_1 := \#_4 (S \cap Q_1), \\
p := \#_4 (Q \cap Q_1), \quad r := \#_4 (S \cap Q \cap Q_1).  
\end{gather*}
With this notation, $\#_{2} S \cap (Q \Delta Q_1) = [t + t_1]_2$. 
Therefore $\#_4 (Q \Delta Q_1) = q + q_1 - 2 p$, and 
\begin{equation*}
\#_4 T_S (Q \Delta Q_1) = 
\begin{cases} 
q + q_1 - 2 p, & \text{if $[t + t_1]_2 = 0$}, \\
- s - (q + q_1 - 2 p) + 2, & \text{if $[t + t_1]_2 = 1$}. 
\end{cases}
\end{equation*}
Taking into account these formulae, one can reduce $c_{Q, Q_1}^{(S)}$ to the following form. 
If $[t]_2 = 0$ and $[t_1]_2 = 0$, then $c_{Q, Q_1}^{(S)} = 0$. 
If $[t]_2 = 1$ and $[t_1]_2 = 0$, then 
$c_{Q, Q_1}^{(S)} = b (q) + b (- q - s + 2 t) + 
c (q + q_1 - 2 p) + c (- s - (q + q_1 - 2 p) + 2)$. 
Similarly, if $[t]_2 = 0$ and $[t_1]_2 = 1$, then 
$c_{Q, Q_1}^{(S)} = b (q_1) + b (- q_1 - s + 2 t_1) + 
c (q + q_1 - 2 p) + c (- s - (q + q_1 - 2 p) + 2)$. 
Finally, if $[t]_2 = 1$ and $[t_1]_2 = 1$, then
$c_{Q, Q_1}^{(S)} = 
b (q) + b (- q - s + 2 t) + 
b (q_1) + b (- q_1 - s + 2 t_1)$. 
The other computations are easier. 

If $\# (Q \cap S)$ is even, i.e. $t = 0, 2$, then 
\begin{gather*} 
c_{Q, \pnt{\xi_0}}^{(S)} = 
b (1) + b (- s + 1) + 
c (q - 1) + c (- s - q + 2 t - 1), \\
c_{Q, \pnt{\eta_0}}^{(S)} = 0, \\
c_{Q, \pnt{\xi_1}}^{(S)} = 
b (1) + b (- s + 1) + c (q + 1) + c (- s - q + 2 t + 1), \\
c_{Q, \pnt{\eta_1}}^{(S)} = 0. 
\end{gather*} 
If $t = 1, 3$, then 
\begin{gather*} 
c_{Q, \pnt{\xi_0}}^{(S)} = 
b (q) + b (- s - q + 2 t) + b (1) + b (- s + 1), \\
c_{Q, \pnt{\eta_0}}^{(S)} = 
b (q) + b (- s - q + 2 t) + c (q - 1) + c (- s - q + 2 t + 1), \\
c_{Q, \pnt{\xi_1}}^{(S)} = 
b (q) + b (- s - q + 2 t) + b (1) + b (- s + 1), \\
c_{Q, \pnt{\eta_1}}^{(S)} = 
b(q) + b (- s - q + 2 t) + c (q + 1) + c (- s - q + 2 t - 1). 
\end{gather*}
There are similar expressions corresponding to $Q_1$. 
If $\# (Q_1 \cap S)$ is even, i.e. $t_1 = 0, 2$, then 
\begin{gather*} 
c_{Q_1, \pnt{\xi_0}}^{(S)} = 
b (1) + b (- s + 1) + 
c (q_1 - 1) + c (- s - q_1 + 2 t_1 - 1), \\
c_{Q_1, \pnt{\eta_0}}^{(S)} = 0, \\
c_{Q_1, \pnt{\xi_2}}^{(S)} = 
b (1) + b (- s + 1) + c (q_1 + 1) + c (- s - q_1 + 2 t_1 + 1), \\
c_{Q_1, \pnt{\eta_2}}^{(S)} = 0. 
\end{gather*}
If $t_1 = 1, 3$, then 
\begin{gather*} 
c_{Q_1, \pnt{\xi_0}}^{(S)} = 
b (q_1) + b (- s - q_1 + 2 t_1) + b (1) + b (- s + 1), \\
c_{Q_1, \pnt{\eta_0}}^{(S)} = 
b (q_1) + b (- s - q_1 + 2 t_1) + c (q_1 - 1) + c (- s - q_1 + 2 t_1 + 1), \\
c_{Q_1, \pnt{\xi_2}}^{(S)} = 
b (q_1) + b (- s - q_1 + 2 t_1) + b (1) + b (- s + 1), \\
c_{Q_1, \pnt{\eta_2}}^{(S)} = 
b(q_1) + b (- s - q_1 + 2 t_1) + c (q_1 + 1) + c (- s - q_1 + 2 t_1 - 1). 
\end{gather*}
Finally, $c_{\pnt{\xi_3}, \pnt{\eta_3}}^{(S)}$ reduces to the form: 
\begin{equation*} 
c_{\pnt{\xi_3}, \pnt{\eta_3}}^{(S)} = 
b (1) + b (- s + 1) + c (2) + c (- s). 
\end{equation*}
For the cardinalities $m_0$, $m_1$, $m_2$, and $m_3$, we have: 
\begin{equation*} 
[m_0]_4 = r, \quad 
[m_1]_4 = t_1 - r, \quad 
[m_2]_4 = t - r, \quad 
[m_3]_4 = s - t - t_1 + r, 
\end{equation*}
where $[ \cdot ]_4$ denotes the canonical image of an integer number in $\mathbb{Z} / 4$. 
Similarly, for the cardinalities $n_0$, $n_1$, $n_2$, and $n_3$, we have: 
\begin{gather*} 
[n_0]_4 = p - r, \quad 
[n_1]_4 = (q_1 - t_1) - (p - r), \quad 
[n_2]_4 = (q - t) - (p - r), \\
[n_3]_4 = (- s) - (q - t) - (q_1 - t_1) + (p - r).  
\end{gather*}
Therefore it remains to investigate what happens to the equation \eqref{c_cond} 
as the parameters $s$, $q$, $q_1$, $t$, $t_1$, $p$, and $r$, vary over $\mathbb{Z} / 4$. 
There are finitely many options, and the corresponding computation is 
easily implemented in Maple. In fact, it is possible to perform it manually, 
if one uses some symmetry of the equation \eqref{c_cond}. 
The result is similar to the case of the equation \eqref{b_cond}, i.e. 
every variant reduces to a linear combination of the 
simple equalities \eqref{cc_bb} mentioned above. 
It means, that we have established the fact that the solvability 
system of equations \eqref{c_cond}, \eqref{b_cond}, 
has solutions, and we have identified at least some of them \eqref{bc_four}. 
 
\section{The orbits} 
 
We are able to construct the group $\mathcal{G}_a$ in two steps. 
First, verify the main condition on $T_c$ for some of the elements of 
$\mathrm{Max} (\mathcal{P}_{T_c} (A_b), \subset)$, and then 
compute the orbits of these elements under the action of $\mathcal{G}_a$. 
One needs enough such elements, so that the orbits cover  
the whole set $\mathrm{Max} (\mathcal{P}_{T_c} (A_b), \subset)$. 
The proof is essentially combinatorial. 

Recall, that for every $U, S \in \mathcal{P} (V)$ 
we have defined the maps $\theta_{S}^{U} : L_b (U) \to L_b (T_S (U))$: 
\begin{equation*} 
\theta_{S}^{U} (\varphi) (v) = \sum_{z \in T_S (\pnt{v})} \varphi (z) + a_{S}^{U} (v), 
\end{equation*}
where $\varphi \in L_b (U)$, $v \in V$. 
There is also a collection of maps $I_{\mu}^{U} : L_b (U) \to L_b (U)$, $U \in \mathcal{P} (V)$, 
corresponding to the gauge transformation with function $\mu : \mathcal{E} (V) \to \mathbb{Z} / 2$, 
defined by the formula
\begin{equation*} 
I_{\mu}^{U} (\varphi) (v) := \varphi (v) + \chi_{\mu}^{U} (v), 
\end{equation*} 
where $\varphi \in L_b (U)$, $v \in V$, and $\chi_{\mu}^{U} (\cdot)$ is as in the previous section. 

Look at the diagram (in $\mathbf{Sets}$): 
\begin{equation*} 
\xymatrix{
L_b (U) \ar@{->}[r] ^{\theta_{S}^{U}} 
\ar@{.>}[d] _{I_{\nu}^{U}} & 
L_b (T_S (U)) \ar@{->}[d] ^{I_{\mu}^{T_S (U)}} \\
L_b (U) \ar@{->}[r] ^{\theta_{S}^{U}} & L_b (T_S (U))
}
\end{equation*}
It turns out, that for every $U, S \in \mathcal{P} (V)$ and every 
$\mu : \mathcal{E} (V) \to \mathbb{Z} / 2$, there exists a unique  
$\nu : \mathcal{E} (V) \to \mathbb{Z} / 2$, rendering this diagram 
commutative. Denote this $\nu$ by $\tau_S (\mu)$. We have
\begin{equation*}
I_{\mu}^{T_S (U)} \circ \theta_{S}^{U} = 
\theta_{S}^{U} \circ I_{\tau_{S} (\mu)}^{U}, 
\end{equation*}
where 
\begin{equation*} 
\tau_S (\mu) (vv_1) := 
\mu (vv_1) + \sum_{z \in T_S (\pnt{v})} \mu (zv_1) + 
\sum_{z_1 \in T_S (\pnt{v_1})} \mu (vz_1), 
\end{equation*}
for $vv_1$ varying over $\mathcal{E} (V)$. 

Now select some sets in $\mathrm{Max} (\mathcal{P}_{T_c} (A_b), \subset)$, and 
verify the main condition for them. 
The most simple case is $M  = \lbrace i_{U}^{b} (\varphi) \rbrace_{\varphi \in L_b (U)}$. 
It is almost obvious, that $M \in \mathrm{Max} (\mathcal{P}_{T_c} (A_b), \subset)$. 
Choose any point in $V$ and denote it by $e$, $e \in V$. 
Put $U = \pnt{e}$. Take any $B \subset M$, and write it as 
$B = \lbrace i_{\pnt{e}}^{b} (\sigma) \rbrace_{\sigma \in S}$, 
$S$ -- some subset of $L_b (\pnt{e})$. 
For $C := M \backslash B$ we have 
$C = \lbrace i_{\pnt{e}}^{b} (\sigma) \rbrace_{\sigma \in S'}$, where 
$S' = L_b (\pnt{e}) \backslash S$. 
It is necessary to show, that if $l \in B^{T_c}$ and $l_1 \in C^{T_c}$, then 
$(l, l_1) \in T_c$. 
We have: $B^{T_c} = C \sqcup (B^{T_c} \backslash C)$ and 
$C^{T_c} = B \sqcup (C^{T_c} \backslash B)$. 
If $l \in C$ or $l_1 \in B$, then the requirement is satisfied. 
The non-trivial case is $l \in B^{T_c} \backslash C$ and 
$l_1 \in C^{T_c} \backslash B$. 
Assume, that such $l$ and $l_1$ exist, and let 
$l = i_{U}^{b} (\varphi)$, $\varphi \in L_b (U)$, and 
$l_1 = i_{U_1}^{b} (\varphi_1)$, $\varphi_1 \in L_b (U_1)$. 
Note, that $U, U_1 \not = \pnt{e}$. Invoking the explicit description \eqref{Tc_def} of 
the relation $T_c$, we conclude, that such $l$ and $l_1$ exist iff 
\begin{gather*} 
\exists \lambda \in \mathbb{Z} / 2 \, 
\forall \sigma \in S \, : \, 
\sum_{v \in \pnt{e} \Delta U} \sigma (v) = \lambda, \\
\exists \lambda' \in \mathbb{Z} / 2 \, 
\forall \sigma \in S' \, : \, 
\sum_{v \in \pnt{e} \Delta U_1} \sigma (v) = \lambda'. 
\end{gather*}
There exist two possibilities: 1) $U = U_1$; 2) $U \not = U_1$. 
Consider the possibility $U = U_1$. 
In this case one must have 
\begin{equation*}
S = \lbrace 
\sigma \in L_b (\lbrace e \rbrace) \, | \, 
\sum_{v \in \pnt{e} \Delta U} \sigma (v) = \lambda
\rbrace, 
\end{equation*}
since otherwise $S'$ cannot satisfy the condition above. 
The parameter $\lambda'$ corresponding to $S'$ is, of course, $\lambda' = 1 + \lambda$. 
For $l = i_{U}^{b} (\varphi)$, using the description of $T_c$, we obtain: 
$\varphi (e) = \lambda + b (\# U) + c (\# (\pnt{e} \Delta U)) + 1$. 
Similarly, for $l_1$ we have: 
$\varphi_1 (e) = \lambda' + b (\# U_1) + c (\# (\pnt{e} \Delta U_1)) + 1 = 
1 + \varphi (e)$. 
Hence, $\varphi_1 (\cdot) \not = \varphi (\cdot)$, and $(l, l_1) \in T_c$. 
Now look at the possibility $U_1 \not = U$. 
This implies that the sets $\pnt{e} \Delta U$ and $\pnt{e} \Delta U_1$ are also different. 
Hence, there exists a point $z$, belonging to one of these sets, and 
not belonging to the other. 
Without loss of generality, let $z \in \pnt{e} \Delta U_1$ and 
$z \not \in \pnt{e} \Delta U$. First, assume, that it is possible to choose them so that 
$z \not = e$. In this case, take any $\sigma$  
such that  $\sum_{v \in \pnt{e} \Delta U_1} \sigma (v) = 1 + \lambda'$. 
Look at $\sum_{v \in \pnt{e} \Delta U} \sigma (v)$. 
If it is equal to $\lambda$, then modify the value of $\sigma (\cdot)$ in the point $z$ by 
adding $1$. 
This does not change the sum with $U_1$, and we obtain 
$\sum_{v \in \pnt{e} \Delta U} \sigma (v) = 1 + \lambda$. 
This $\sigma$ belongs neither to $S$, nor to $S'$. But this is a contradiction, 
since $S$ and $S'$ partition the set $L_b (\pnt{e})$ of all possible $\sigma$. 
Therefore, the pair $(l, l_1)$ cannot exist. 
It remains to consider the case when the only option for $z$ is $z = e$. 
We have: $U \not \ni e$ and $U_1 = \pnt{e} \sqcup U$. 
Then the parameters $\lambda$ and $\lambda'$ associated to $S$ and $S'$ may be written as 
$\lambda = b (1) + \sum_{v \in U} \sigma (v)$, $\sigma$ -- any element of $S$, and 
$\lambda' = \sum_{v \in U} \sigma' (v)$, $\sigma'$ -- any element of $S'$. 
Since $S$ and $S'$ partition $L (\pnt{e})$, $S'$ has to coincide with the set of \emph{all} 
$\sigma'$ such that $\sum_{v \in U} \sigma' (v) = \lambda'$ (otherwise it is impossible 
to define $\lambda$ for $S$). Therefore, for every $\sigma \in S$ we have 
$\sum_{v \in U} \sigma (v) = 1 + \lambda'$, and one obtains 
$\lambda = b (1) + 1 + \lambda'$. 
Since $l = i_{U}^{b} (\varphi)$ is in relation $T_c$ with every $i_{\pnt{e}}^{v} (\sigma)$, 
invoking the definition of $L (U)$ and the description of $T_c$, it follows that: 
$\varphi (e) + b (\#_4 U) + \lambda = c (\#_4 U + 1) + 1$. 
Similarly, for $\varphi_1 \in L (\pnt{e} \Delta U)$, we arrive at: 
$\varphi_1 (e) + b (\#_4 U + 1) + \big[ 
1 + \lambda + b (1) \big] = c (\#_4 U) + 1$. Hence, 
\begin{equation*} 
\varphi (e) + \varphi_1 (e) = 
1 + b (1) + b (\#_4 U) + b (\#_4 U + 1) + c (\#_4 U) + c (\#_4 U + 1). 
\end{equation*}
On the other hand, the requirement $(i_{U}^{b} (\varphi), i_{U_1}^{b} (\varphi_1))$ implies, that 
$\varphi (e) + \varphi_1 (e) = 1 + c (1)$. Therefore, one obtains a condition 
\begin{equation*} 
b (1) + c (1) + 
b (\#_4 U) + c (\#_4 U) + 
b (\#_4 U + 1) + c (\#_4 U + 1) = 0. 
\end{equation*}
Since this has to be valid for generic $U$, we obtain: 
\begin{equation}
\begin{gathered} 
b (0) + c (0) = 0, \quad b (2) + c (2) = 0, \\
b (1) + c (1) + b (3) + c (3) = 0. 
\end{gathered}
\label{bc0bc2bc13}
\end{equation}
The latter is the equation we already have, and the first two imply 
the other equation, but are not equivalent to it. 
Hence, under these conditions, the main property of $T_c$ for the set 
$M = \lbrace i_{\pnt{e}}^{b} (\sigma) \rbrace_{\sigma \in L_b (\pnt{e})}$ is established.  

Let us consider some other subsets 
$M \in \mathrm{Max} (\mathcal{P}_{T_c} (A_b), \subset)$. 
There exists a natural map $\eta : A_b \to \mathcal{P} (V)$, 
$i_{U}^{b} (\varphi) \mapsto U$. For every $B \subset A_b$, call the set 
$\lbrace \eta (l) \rbrace_{l \in B}$ the \emph{shadow} of $B$. 
Take any non-empty subset $\Omega \subset V$. 
Under some additional assumptions on $b (\cdot)$ and $c (\cdot)$, it will be shown that 
there exist sets $M \in \mathrm{Max} (\mathcal{P}_{T_c} (A_b), \subset)$ of the form 
\begin{equation*} 
M = \bigsqcup_{U \in \mathcal{P}_{\mathrm{odd}} (\Omega)} 
\lbrace i_{U}^{b} (\varphi) \rbrace_{\varphi \in Q_U}, 
\end{equation*} 
where $Q_{U}$ are some subsets of $L_b (U)$, and 
\begin{equation*} 
\mathcal{P}_{\mathrm{odd}} (\Omega) := \lbrace 
U \subset \Omega \, | \, \text{$\# U$ is odd} \rbrace. 
\end{equation*}
Similarly, one may introduce the set 
$\mathcal{P}_{\mathrm{even}} (\Omega)$ consisting of all subsets of $\Omega$ of even cardinality. 
We will impose such conditions of $b (\cdot)$ and $c (\cdot)$, that 
the following statement will be true: 
if $B \subset A_b$ has a shadow which contains a subset 
being an element of $\mathcal{P}_{\mathrm{even}} (\Omega)$, 
then it does not belong to $\mathcal{P}_{T_c} (A_b)$. 

More precisely, take any $\Omega \subset V$, such that $\# \Omega$ is even. 
Assume that $b (\cdot)$ and $c (\cdot)$ satisfy the conditions \eqref{bc0bc2bc13}. 
Is it possible to have a set $B \in \mathcal{P}_{T_c} (A_b)$ consisting of 
$\# \Omega + 1$ elements, such that $\# \Omega$ of them are of the form  
$i_{\pnt{v}}^{b} (\sigma_v)$, $\sigma_v \in L_b (\pnt{v})$, $v \in \Omega$, and 
and the other element is of the form $i_{\Omega}^{b} (\varphi)$, $\varphi \in L_b (\Omega)$?
Denote $\mathcal{E} (\Omega) := \lbrace U \subset \Omega \, | \, \# U = 2 \rbrace$. 
Assume that $i_{\pnt{v}}^{b} (\sigma_v)$, $\sigma_v \in L_b (\pnt{v})$, $v \in \Omega$, 
are pairwise in relation $T_c$. 
For $z \not = v$ we have: 
$\sigma_z (v) + \sigma_v (z) = 1 + c (2)$. 
Choose and fix any order $\prec$ on $V$ and 
associate to this collection of elements a 
function $\tau : \mathcal{E} (\Omega) \to 
\mathbb{Z} / 2$, $\tau (zw) := \sigma_v (z)$, $v \prec z$. 
Hence, for any $vv_1 \in \mathcal{E} (\Omega)$, 
\begin{equation*} 
\sigma_{v} (v_1) = 
\begin{cases} 
\tau (vv_1), & \text{if $v \prec v_1$}, \\
\tau (vv_1) + 1 + c (2), & \text{if $v \succ v_1$}. \\
\end{cases}
\end{equation*}  
Now investigate what this means for $\varphi$.
For every $v \in \Omega$, the definition of $T_c$ yields: 
\begin{equation*} 
\sum_{z \in \pnt{v} \Delta \Omega} ( \sigma_{v} (z) + \varphi (z) ) = 
c (\#_4 \Omega - 1) + 1.  
\end{equation*}
The fact $\sum_{z \in \Omega} \varphi (z) = b (\#_4 \Omega)$ yields: 
\begin{equation*} 
\varphi (v) = \sum_{z \in \Omega \backslash \pnt{v}} 
\sigma_{v} (z) + b (\#_4 \Omega) + c (\#_4 \Omega - 1) + 1,  
\end{equation*}
where $v \in \Omega$. 
Apply summation over $v \in \Omega$ and invoke once more the mentioned fact to obtain:  
\begin{equation*} 
\frac{m (m - 1)}{2} \, \big( c (2) + 1 \big) + 
m \big[ 
b (m) + c (m - 1) + 1
\big] = b (m). 
\end{equation*}
where $m := \#_4 \Omega$.  
If this were true for generic $\Omega$, one would have the following four 
equalities corresponding to $m = 0, 1, 2, 3$ respectively: 
$b (0) = 0$, $c (0) + 1 = 0$, $(c (2) + 1) + b (2) = 0$, and 
$(c (2) + 1) + c (2) + 1 = 0$. 
The latter is just an identity. 
The third one is not valid, since we already have a condition 
$b (2) + c (2) = 0$. 
Moreover, since $b (0) + c (0) = 0$, either the first or the second 
equality is not valid as well. Put $b (0) = c (0) = 1$. 
Hence, $m$ cannot be $0$ or $2$, i.e. $\# \Omega$ cannot be even. 
We have 
\begin{equation}
\begin{gathered} 
b (0) = 1, \quad c (0) = 1, \\
b (2) + c (2) = 0, \\
b (1) + c (1) + b (3) + c (3) = 0. 
\end{gathered}
\label{b0c0bc2bc13}
\end{equation}
It is impossible to have a collection consisting of elements of $A_b$ of the form 
$i_{v}^{b} (\sigma_v)$, $v \in \Omega$, and $i_{\Omega}^{b} (\varphi)$, if 
$\# \Omega$ is even. 
In case $\# \Omega$ is odd, the values of $\varphi (\cdot)$ are determined by the 
function $\tau : \mathcal{E} (\Omega) \to \mathbb{Z} / 2$, 
associated to $\sigma_v (\cdot)$, $v \in \Omega$.  
The values on the points of $V \backslash \Omega$ can be chosen arbitrary. 

Now suppose one has a collection of elements $l_1, l_2, \dots, l_n \in A_b$ 
which are pairwise in relation $T_c$. Denote $U_i := \eta (l_i)$, $i = 1, 2, \dots, m$, 
where $\eta : A_b \to \mathcal{P} (V)$ is the natural map mentioned above. 
Some of these sets may have cardinality $1$, and some may contain more points. 
Denote by $\Omega$ the union of all $U_i$ such that $\# U_i = 1$. 
Note, that it is possible that $\Omega$ is empty. 
There exists a bijection $A_b \os{\sim}{\to} A_b$ which respects the relation $T_c$, 
which transforms this collection into a collection 
with the following property: every $U_i$ is a subset of $\Omega$. 
Indeed, we have constructed the maps $\wh{\theta}_{S}^{(\ov{a})}$. 
These maps respect the relation $T_c$ on $A_b$. 
If $l \in A_b$ satisfies $\eta (l) = U$, then $l' = \wh{\theta}_{S}^{(\ov{a})} (l)$ satisfies 
$\eta (l') = T_S (U)$. 
Let $l_1, l_2, \dots, l_n \in A_b$ be as above. 
Denote $U_i := \eta (l_i)$, $i = 1, 2, \dots, n$, and 
construct the corresponding $\Omega$. If there exists $U_{i_0}$, which is not a subset of $\Omega$, 
then one can take a point $v_0 \in U_0 \backslash \Omega$. 
Look at the composition $T_{\ov{U}_{i_0}} \circ T_{ \ov{U}_{i_0} \sqcup \pnt{v_0} } \circ 
T_{\ov{U}_{i_0}}$. This map transfers $U_{i_0}$ into a one-point set $\pnt{v_0}$, and at the same time 
leaves all the one point-sets $\lbrace v \rbrace$, $v \in \Omega$, fixed. 
Therefore, if one applies a composition 
$\wh{\theta}_{\ov{U}_{i_0}}^{(a)} \circ 
\wh{\theta}_{ \ov{U}_{i_0} \sqcup \pnt{v_0} }^{(a)} \circ 
\wh{\theta}_{\ov{U}_{i_0}}^{(a)}$ to each $l_1, l_2, \dots, l_n$, 
one increases the number of points in $\Omega$ by $1$. 
Proceeding this way we arrive at the situation where  
all $U_i$ are subsets of the corresponding $\Omega$. 
Of course, in this case, all $U_i$ will have odd cardinalities. 
Note, that the cardinality of $\Omega$ need not be odd. 

Take any $l_1, l_2, \dots, l_n$, such that 
$\eta (l_i) = \pnt{e_i}$, $i =  1, 2, \dots, n$, 
$e_i \in V$ some points, such that 
$e_i \not = e_j$ for $i \not = j$. 
Assume that $(l_i, l_j) \in T_c$, $i \not = j$. 
Hence $\Omega = \lbrace e_1, e_2, \dots, e_n \rbrace$. 
Take any $U \subset \Omega$ and try to construct 
$l \in A_b$ of the form $l = i_{U}^{b} (\varphi)$, $\varphi \in L_b (U)$, such that for all $i$, 
$(l, l_i) \in T_c$. The cardinality  $\# U$ needs to be odd. 
Let $U = \lbrace e_i \rbrace_{i \in I}$, where 
$I \subset \lbrace 1, 2, \dots, n \rbrace$, $\# I$ is odd. 
The elements $l_k$, $k = 1, 2, \dots, n$, 
are of the form $l_k = i_{\pnt{e_k}}^{b} (\sigma_k)$, the $\sigma_k$ 
element of $L_b (\pnt{e_k})$. 
The requirement that $(l_i, l) \in T_c$ for every $i \in I$, yields: 
\begin{equation*} 
\varphi (e_i) = 
\sum_{i' \in I \backslash \pnt{i}} 
\sigma_{i} (e_{i'}) + b (\#_4 I) + c (\#_4 I - 1) + 1.  
\end{equation*} 
Similarly, the requirement that for every $q \in \lbrace 1, 2, \dots, n \rbrace \backslash I$, the pair 
$(l_q, l) \in T_c$, yields: 
\begin{equation*} 
\varphi (e_q) = 
\sum_{i \in I} \sigma_q (e_i) + b (1) + b (\#_4 I) + c (\#_4 I + 1) + 1. 
\end{equation*}
Therefore, the values of $\varphi (\cdot)$ on the points of $\Omega$ 
are determined, and on the points of $V \backslash \Omega$ remain arbitrary. 
Now take any $W \subset \Omega$, $W \not = U$, $\# W$ is odd. 
Let $W = \lbrace e_j \rbrace_{j \in J}$, $J \subset \lbrace 1, 2, \dots, n \rbrace$. 
There exists $l' \in A_b$ of the form $l' = i_{W}^{b} (\psi)$, $\psi \in L_b (W)$, 
which is in relation $T_c$ with every $l_1, l_2, \dots, l_n$. 
The values of the function $\psi (\cdot)$ on the points of $\Omega$ are given by formulae similar 
to the ones above, and on $V \backslash \Omega$ can be assigned in an arbitrary way. 
Is it possible to have $(l', l) \in T_c$? 
It turns out, that $l$ and $l'$ are \emph{always} in $T_c$.   
Note, that the condition for $(l, l') \in T_c$ involves only the values of 
$\varphi (\cdot)$ and $\psi (\cdot)$ in the points of $\Omega$ (more precisely, only in 
$e_s$, $s \in I \Delta J$). For these values one has the corresponding expressions via 
$\sigma_k (\cdot)$, $k = 1, 2, \dots, n$. 
Substitute them into the mentioned condition and
take into account, that $\sigma_{k} (e_{k'}) + \sigma_{k'} (e_k) = c (2) + 1$. 
After simplification, the expression reduces to: 
\begin{multline*} 
\frac{\# (I \Delta J) (\# (I \Delta J) - 1)}{2} 
\big( c (2) + 1 \big) 
+ c (\#_4 (I \Delta J)) + 1 + \\ +  
\# (I \cap \ov{J}) \big\lbrace 
c (\#_4 I - 1) + c (\#_4 J + 1)
\big\rbrace + \\ + 
\# (J \cap \ov{I}) \big\lbrace 
c (\#_4 J - 1) + c (\#_4 I + 1)
\big\rbrace +
\\ + 
\# (I \Delta J) \big[ 
b (1) + b (\#_4 I) + b (\#_4 J)
\big] = 0. 
\end{multline*}
In order to compute the value of the left-hand side it suffices to know 
$\#_4 I$, $\#_4 J$, and $\#_4 (I \cap J)$. 
Recall, that $\#I$ and $\# J$ are odd. 
Hence, it remains to run through all the $2 \times 2 \times 4 = 16$ 
(in fact, even $8$, due to the symmetry with respect to permutation of $I$ and $J$)
possibilities and look at 
what happens to the equation above. 
A straightforward (Maple) computation shows that each 
time one obtains either an identity $0 \equiv 0$, or an equality 
$c (0) = 1$. The latter is already present in the list of assumptions \eqref{b0c0bc2bc13} 
concerning $b (\cdot)$ and $c (\cdot)$ above. 
Therefore, indeed $(l, l') \in T_c$. 
 
Associate to the set $l_1, l_2, \dots, l_n$ the function 
$\tau : \mathcal{E} (\Omega) \to \mathbb{Z} / 2 $ as explained above. 
One may write all formulae in terms of this function. 
Construct from it a function 
$\wh{\tau} : \Omega \times \Omega \to \mathbb{Z} / 2$ 
of two arguments, 
\begin{equation} 
\wh{\tau} (v, v_1) := 
\begin{cases} 
\tau (vv_1), & \text{if $v \prec v_1$}, \\
b (1), & \text{if $v = v_1$}, \\
\tau (vv_1) + c (2) + 1, & \text{if $v \succ v_1$}. 
\end{cases}
\label{tau_hat}
\end{equation}
Note, that for $v \not = v_1$, one has  
$\wh{\tau} (v, v_1) = \wh{\tau} (v_1, v) + c (2) + 1$. 
Next, construct a function 
$\wt{\tau} : \mathcal{P}_{\mathrm{odd}} (\Omega) \times \Omega \to \mathbb{Z} / 2$, as follows. 
Put $\wt{\tau} (\pnt{v}, v) := \wh{\tau} (v, v) = b (1)$, and for $U \not = \pnt{v}$ put 
\begin{equation} 
\wt{\tau} (U, v) := 
\sum_{z \in U} \wh{\tau} (v, z) + 
b (\#_4 U) + b (1) + c (\#_4 (U \Delta \pnt{v})) + 1. 
\label{tau_tilde}
\end{equation}
It is convenient to rewrite the formulae obtained above using this notation. 
For $l_i = i_{\pnt{e_i}}^{b} (\sigma_i)$, $i = 1, 2, \dots, n$, 
we have: 
\begin{equation*} 
\forall z \in \Omega: \sigma_i (z) = \wt{\tau} (\pnt{e_i}, z). 
\end{equation*} 
 For an element $l$ of the form $l = i_{U}^{b} (\varphi)$, $\varphi \in L_b (U)$, 
$U \subset \Omega$, $\# U$ is odd, 
which is in relation $T_c$ with every $l_i$, we have: 
\begin{equation*} 
\forall z \in \Omega: \varphi (z) = \wt{\tau} (U, z). 
\end{equation*}
The values of $\sigma_i (w)$ and $\varphi (w)$ in $w \in V \backslash \Omega$ remain arbitrary. 

Now, take any non-empty $\Omega \subset V$, and take any function 
$\tau : \mathcal{E} (\Omega) \to \mathbb{Z} / 2$. 
Define $\wt{\tau}$ corresponding to $\tau$ by the formulae \eqref{tau_hat}, \eqref{tau_tilde}. 
For every $U \in \mathcal{P}_{\mathrm{odd}} (\Omega)$, denote 
\begin{equation} 
Q_{U} := \lbrace 
\varphi \in L_b (U) \, | \, 
\forall z \in \Omega : \varphi (z) = \wt{\tau} (U, z)
\rbrace. 
\label{QUset}
\end{equation} 
Note, that every $\varphi \in Q_{U}$ should satisfy 
$\sum_{z \in U} \wt{\tau} (U, z) = b (\#_4 U)$. 
This yields the following condition: 
\begin{equation*}
\frac{\# U (\# U - 1)}{2} \, \big( c (2) - 1 \big)+ 
\# U c (\#_4 U - 1) - 1 = 0.  
\end{equation*} 
The value of the left-hand side is determined by $\#_4 U$. 
Since $\# U$ is odd, it is necessary to consider just two cases: 
$\#_4 U = 1$ and $\#_4 U = 3$. 
In the first case one obtains $c (0) = 1$, i.e. the condition we already have above, 
and the second case reduces to $0 \equiv 0$.  

Consider now the following set (for some $\Omega$ and $\tau$): 
\begin{equation} 
M := \bigsqcup_{U \in \mathcal{P}_{\mathrm{odd}} (\Omega)} 
\lbrace i_{U}^{b} (\varphi) \rbrace_{\varphi \in Q_U}. 
\label{Mset}
\end{equation}
The elements of $M$ are pairwise in relation $T_c$. 
The cardinality of $\Omega$ is $n$, and the cardinality of $V$ is $N$. 
On the points of $V \backslash \Omega$ a function $\varphi \in Q_U$ may take any value. 
In total there are $2^{N - n}$ possibilities for that. 
The number of all subsets of $\Omega$ is $2^{n}$, and among them the number of those with odd 
cardinality is $2^{n - 1}$. Hence $\# M = 2^{n - 1} \times 2^{N - n} = 2^{N - 1}$. 
This number coincides with $\# L_b (W)$ for every non-empty $W \subset V$. 
It is not difficult to show that $M \in \mathrm{Max} (\mathcal{P}_{T_c} (A_b), \subset)$. 
Indeed, if $\Omega = V$, then it is impossible to have an element of the form 
$i_{W}^{b} (\psi)$ which is in relation $T_c$ with every element of $M$, since if 
$\# W$ is odd, then such element is already in $M$, and the case $\# W$ being even is 
excluded due to conditions above. 
Consider a proper non-empty $\Omega$. 
For same reasons, $W$ cannot be a subset of $\Omega$. 
Hence, there exists a point $w \in W \backslash \Omega$. 
Consider $\theta' := \theta_{\ov{W}}^{(\ov{a})} \circ 
\theta_{\ov{W} \sqcup \pnt{w}}^{(\ov{a})} \circ 
\theta_{\ov{W}}^{(\ov{a})}$. Apply $\theta'$ to every element of $M$ and to $i_{W}^{b} (\psi)$. 
The set $M$ will still be of the form as above, but, perhaps, corresponding to a different $\tau$, 
and the image of $i_{W}^{b} (\psi)$ after $\theta'$ is projected by the natural map 
$\eta : A_{b} \to \mathcal{P} (V)$ into the point $\pnt{w}$. 
Therefore, it suffices to consider just the case $W = \lbrace w \rbrace$. 
Take any $z \in \Omega$ and look at $\lbrace i_{\pnt{z}}^{b} (\varphi) \rbrace_{\varphi \in Q_{\pnt{z}}}$. 
As $\varphi$ varies over $Q_{\pnt{z}}$, its value in $w$ sweeps up the whole $\mathbb{Z} / 2$. 
Therefore, it is impossible to satisfy $(i_{\pnt{z}}^{b} (\varphi), i_{\pnt{w}}^{b} (\psi) ) \in T_c$ 
simultaneously for all $\varphi$. Hence, $M$ is maximal. 

Choose and fix any $B \subset M$. 
One has: 
\begin{equation}
B = \bigsqcup_{U \in \mathcal{P}_{\mathrm{odd}} (\Omega)} 
\lbrace i_{U}^{b} (\varphi) \rbrace_{\varphi \in S_U}, 
\label{Bset}
\end{equation}
where $S_{U} \subset Q_U$ are subsets, (some of $S_U$, or even all, may be empty). 
Take any $i_{W}^{b} (\psi)$, $W \in \mathcal{P} (V)$, $\psi \in L_b (W)$, and 
look at what the condition $i_{W}^{b} (\psi) \in B^{T_c}$ means. 
It is necessary to consider different possibilities for $W$. 
Start with the case where $W$ is a subset of $\Omega$, and the number of elements in it is even. 
For any $U \in \mathcal{P}_{\mathrm{odd}} (\Omega)$ and any $\varphi \in S_{U}$, one must have 
$\sum_{z \in U \Delta W} (\psi (z) + \varphi (z)) = c (\#_4 (U \Delta W)) + 1$. 
This is equivalent to: 
\begin{equation*}
\sum_{z \in U} \psi (z) =  
\sum_{z \in W} \varphi (z) + 
b (\#_4 W) + b (\#_4 U) + c (\#_4 (U \Delta W)) + 1.  
\end{equation*}
Since $W \subset \Omega$, the values of $\varphi (z)$ are known: 
$\varphi (z) = \wt{\tau} (U, z)$. 
Therefore, in the case $W \in \mathcal{P}_{\mathrm{even}} (\Omega)$, the 
requirement $i_{W}^{b} (\psi) \in B^{T_c}$ is rewritten as follows: 
\begin{multline} 
\big( i_{W}^{b} (\psi) \in B^{T_c} \big) = 
\bigwedge_{U \in \mathcal{P}_{\mathrm{odd}} (\Omega)} 
\Big( 
\text{$S_{U} = \emptyset$ or  
$\sum_{z \in U} \psi (z) =  
\sum_{z \in W}  \wt{\tau} (U, z) +$} 
\\ \text{$+ b (\#_4 W) + b (\#_4 U) + c (\#_4 (U \Delta W)) + 1$}
\Big).
\label{Weven}
\end{multline}
There is a similar expression in case $W \in \mathcal{P}_{\mathrm{odd}} (\Omega)$, 
but special care is needed for the variant $U = W$: 
\begin{multline} 
\big( i_{W}^{b} (\psi) \in B^{T_c} \big) = 
\big( \psi \not \in S_{W} \big) \, \& \, 
\bigwedge_{\substack{ U \in \mathcal{P}_{\mathrm{odd}} (\Omega), \\ U \not = W}}  
\Big( 
\text{$S_{U} = \emptyset$ or  
$\sum_{z \in U} \psi (z) =$} 
\\ \text{$= \sum_{z \in W}  \wt{\tau} (U, z) + b (\#_4 W) + b (\#_4 U) + c (\#_4 (U \Delta W)) + 1$}
\Big).
\label{Wodd}
\end{multline}
Finally, there is a more complicated case, when $W$ contains a part outside $\Omega$, i.e. 
$W \cap \ov{\Omega} \not = \emptyset$. 
In this case we have to deal with the sum $\sum_{z \in W} \varphi (z)$, 
but only part of $\varphi (z)$ are known, i.e. those that correspond to $z \in W \cap \Omega$, 
can be expressed as $\wt{\tau} (U, z)$. 
Instead of the equality above, we obtain: 
\begin{multline} 
\big( i_{W}^{b} (\psi) \in B^{T_c} \big) = 
\bigwedge_{U \in \mathcal{P}_{\mathrm{odd}} (\Omega)} 
\Big( 
\text{$S_{U} = \emptyset$ or  
$\sum_{z \in U} \psi (z) = $}\\
\text{$\sum_{z \in W \cap \Omega}  \wt{\tau} (U, z) +
\sum_{z \in W \cap \ov{\Omega}} 
\big( \text{any $\varphi \in S_U$}\big) (z)$} +  
\\ \text{$+ b (\#_4 W) + b (\#_4 U) + c (\#_4 (U \Delta W)) + 1$}
\Big).
\label{Wbeyond}
\end{multline}
The sum on the right-hand side containing $\varphi$ should not depend on the choice of $\varphi \in S_{U}$, 
so we have a condition on $S_{U}$ ensuring the existence of $i_{W}^{b} (\psi) \in B^{T_c}$ with such $W$. 

First, consider in more detail the 
term $\psi \not \in S_{W}$, $W \in \mathcal{P}_{\mathrm{odd}} (\Omega)$ in the formula \eqref{Wodd}. 
Since $S_{W} \subset Q_{W}$, it splits into a disjunction 
$(\psi \in Q_{W} \backslash S_{W}) \vee (\psi \not \in Q_{W})$. 
The case that is described by the second term means that 
there exists a non-trivial $\chi : \Omega \to \mathbb{Z} / 2$, such that 
$\psi (z) = \wt{\tau} (W, z) + \chi (z)$, $z \in \Omega$. 
Since $\psi \in L_b (W)$, $\chi$ must satisfy $\sum_{z \in W} \chi (z) = 0$. 
It is convenient to view $\chi$ as an indicator function $\chi_{Z}$ of some 
non-empty subset $Z \in \mathcal{P} (\Omega)^{\times} := \mathcal{P} (\Omega) \backslash 
\lbrace \emptyset \rbrace$. 
Therefore, we have 
\begin{equation*} 
(\psi \not \in S_{W}) = (\psi \in Q_{W} \backslash S_{W}) \vee (\psi \not \in Q_{W}), 
\end{equation*} 
while  
\begin{equation*} 
(\psi \not \in Q_{W}) = 
\bigvee_{\substack{Z \in \mathcal{P} (\Omega)^{\times}, \\ \#_2 (Z \cap W) = 0}} 
\Big[ 
\psi (z \in \Omega) = \wt{\tau} (W, z) + \chi_{Z} (z) 
\Big]. 
\end{equation*}
Note that the cardinality of $Z$ need not be odd. 

The expression for $i_{W}^{b} (\psi) \in B^{T_c}$, $W \in \mathcal{P}_{\mathrm{odd}} (\Omega)$, 
reduces to: 
\begin{multline} 
\big( i_{W}^{b} (\psi) \in B^{T_c} \big) = 
(\psi \in Q_{W} \backslash S_{W}) \text{\, or \,} 
\bigvee_{\substack{Z \in \mathcal{P} (\Omega)^{\times}, \\ \#_2 (Z \cap W) = 0}} 
\Big\lbrace  
\big[ \psi (z \in \Omega) = \\ = \wt{\tau} (W, z) + \chi_{Z} (z) \big]
\, \& \,  
\bigwedge_{
\substack{
U \in \mathcal{P}_{\mathrm{odd}} (\Omega), \\ 
\#_2 (U \cap Z) = 1
}}
(S_U = \emptyset)
\Big\rbrace. 
\label{Wodd_reduced}
\end{multline}
Note, that for the conjunction on the right-hand side of the formula one first obtains the 
range of possible values of $U$ in the form 
$U \in \mathcal{P}_{\mathrm{odd}} (\Omega)$,  
$\#_2 ((U \Delta W) \cap Z) = 1$, but since 
$\#_2 (Z \cap W) = 0$, one has $\#_2 ((U \Delta W) \cap Z) = \#_2 (U \cap Z)$. 

Now one needs to consider arbitrary $i_{W}^{b} (\psi) \in B^{T_c}$ and 
$i_{W_1}^{b} (\psi_1) \in (M \backslash B)^{T_c}$, and then 
verify that $i_{W_1}^{b} (\psi_1)$ is in relation $T_c$ with $i_{W}^{b} (\psi)$. 
The formulae for $i_{W_1}^{b} (\psi_1)$ are similar, except that it is necessary to replace 
all $S_U$ with $Q_U \backslash S_U$.
We have to establish the following implication: 
\begin{equation} 
\big( i_{W}^{b} (\psi) \in B^{T_c} \big) \, \& \, 
\big( i_{W_1}^{b} (\psi_1) \in (M \backslash B)^{T_c} \big) 
\Rightarrow 
(i_{W}^{b} (\psi), i_{W_1}^{b} (\psi_1)) \in T_c. 
\label{WW1_implic}
\end{equation} 
There are three possibilities for $W$ and three possibilities for $W_1$ described above. 
In total, due to the symmetry of $T_c$, this yields $3 + 3 (3 - 1) / 2 = 6$ combinations. 
Each needs to be investigated separately verifying whether a  
strengthening of the conditions on $b (\cdot)$ and $c (\cdot)$ results.  
The result is the following. 
\begin{thm}
Assume that the number $N$ of points in $V$ is divisible by $4$. 
Let $b, c : \mathbb{Z} /4 \to \mathbb{Z} / 2$ be two functions, such that 
$b (0) = c (0) = 1$, $b (2) + c (2) = 0$, and 
$\sum_{i = 1, 3} (b (i) + c (i)) = 0$. 
Define the finite set $A_b$ and the relation $T_c$ on $A_b$ by \eqref{Ab_def}, \eqref{Tc_def} 
using \eqref{bc_four}. 
Then the relation $T_c$ is saturated. 
 
\end{thm}
\emph{Proof}. 
The aim is to establish the implication \eqref{WW1_implic}.  
The proof splits naturally into six parts, corresponding to 
the six combinations mentioned above. 

1) The case $W, W_1 \in \mathcal{P}_{\mathrm{odd}} (\Omega)$. 
We have a formula \eqref{Wodd} for $W$, and there exists a similar formula for $W_1$ 
obtained after replacing $S_U$ by $Q_U \backslash S_U$. 
If $W = W_1$, then one needs to show that $\psi (\cdot) \not = \psi_1 (\cdot)$. 
In this case, for some $Z \in \mathcal{P} (\Omega)^{\times}$,  
$\psi (z \in \Omega) = \wt{\tau} (W, z) + \chi_{Z} (z)$, and 
for all $U \in \mathcal{P}_{\mathrm{odd}} (\Omega)$ such that $\# ((U \Delta W) \cap Z)$ is odd, 
$S_U = \emptyset$. Similarly, for some $Z_1 \in \mathcal{P} (\Omega)^{\times}$,  
$\psi_1 (z \in \Omega) = \wt{\tau} (W, z) + \chi_{Z_1} (z)$, and 
for all $U_1 \in \mathcal{P}_{\mathrm{odd}} (\Omega)$ such that $\# ((U_1 \Delta W) \cap Z_1)$ is odd, 
$Q_{U_1} \backslash S_{U_1} = \emptyset$. 
If $Z \not = Z_1$, then $\psi (\cdot) \not = \psi_1 (\cdot)$, since 
$\chi_{Z} (\cdot) \not = \chi_{Z_1} (\cdot)$. Hence, the implication holds. 
If $Z = Z_1$, then for every of the mentioned $U$, we have $S_{U} = \emptyset$ and 
$Q_{U} \backslash S_{U} = \emptyset$. Since $Q_{U}$ is not empty, this possibility cannot occur. 
So the implication is established for $W = W_1$. 
Now assume, that $W \not = W_1$. 
One needs to verify that 
$\sum_{z \in W \Delta W_1} \big( \psi (z) + \psi_1 (z) \big) = c (\#_4 (W \Delta W_1)) + 1$. 
Observe that from the definition of $M$, and the property that relates $Q_U$ and $\wt{\tau} (U, z)$, 
for any $U, U_1 \in \mathcal{P}_{\mathrm{odd}} (\Omega)$, we have 
\begin{equation*} 
\sum_{z \in U \Delta U_1} (\wt{\tau} (U, z) + \wt{\tau} (U_1, z)) = c (\#_4 (U \Delta U_1)) + 1. 
\end{equation*}
Hence, if $\psi \in Q_{W} \backslash S_{W} \subset Q_{W}$ and $\psi_1 \in S_{W_1} \subset Q_{W_1}$, 
the requirement is satisfied. 
Now let $\psi \not \in Q_{W}$, but $\psi_1 \in S_{W_1}$. 
For $\psi (\cdot)$ we have a non-empty $Z \subset \Omega$, such that 
$\psi (z \in \Omega) = \wt{\tau} (W, z) + \chi_{Z} (z)$. 
Moreover, whenever $U \in \mathcal{P}_{\mathrm{odd}} (\Omega)$ and 
$\# ((U \Delta W) \cap Z)$ is odd, one has $S_{U} = \emptyset$. 
For $\psi_1 (\cdot)$ we have $\psi_1 (z \in \Omega) = \wt{\tau} (W, z)$. 
Therefore, the required equality holds iff $\sum_{z \in W \Delta W_1} \chi_{Z} (z) = 0$, i.e. 
$\# ((W \Delta W_1) \cap Z)$ is even. But $\# ((W \Delta W_1) \cap Z)$ cannot be odd, since 
then (specializing $U$ to $W_1$) one obtains $S_{W_1} = \emptyset$, i.e. $\psi_1 (\cdot)$ does not exist. 
Hence, in this case the implication is established. 
The dual case, i.e. $\psi \in Q_{W} \backslash S_{W}$ and $\psi_1 \not \in  Q_{W_1}$, 
is completely similar. 
It remains to investigate the possibility $\psi \not \in Q_{W}$ and $\psi_1 \not \in Q_{W_1}$. 
For some non-empty $Z \subset \Omega$, $\# (Z \cap W)$ even, one has 
$\psi (z \in \Omega) = \wt{\tau} (W, z) + \chi_{Z} (z)$. 
Similarly, for some non-empty $Z_1 \subset \Omega$, $\# (Z_1 \cap W_1)$ even, one has 
$\psi_1 (z \in \Omega) = \wt{\tau} (W_1, z) + \chi_{Z_1} (z)$. 
One needs an equality $\sum_{z \in W \Delta W_1} (\chi_Z (z) + \chi_{Z_1} (z)) = 0$, i.e. 
$\# ((W \Delta W_1) \cap Z)$ and $\# ((W \Delta W_1) \cap Z_1)$ are either both odd, or both even. 
To establish it, use the conjunctions over $U$ and $U_1$ present in the corresponding formulae. 
Note, that $Z \cap Z_1$ needs to be empty. 
Indeed, otherwise one may take any $v \in Z \cap Z_1$ and put $U = U_1 = \pnt{v}$. 
Since for these $U$ and $U_1$, $\#_2 (U \cap Z) = \#_2 (U_1 \cap Z_1) = 1$, we have  
$S_{\pnt{v}} = \emptyset$ and $S_{\pnt{v}} = Q_{\pnt{v}}$, this  
contradicts $Q_{\pnt{v}} \not = \emptyset$. 
Hence, $Z \cap Z_1 = \emptyset$. 
Moreover, $Z$ and $Z_1$ should partition $\Omega$, since otherwise one can put 
$U = U_1 = \lbrace v, v_1, w \rbrace$, where $v \in Z$, $v_1 \in Z_1$, and $w \in \Omega \backslash 
(Z \sqcup Z_1)$. For these $U$ and $U_1$ again have intersections with $Z$ and $Z_1$, respectively, of odd 
cardinalities, and one obtains a contradiction between 
$S_{\lbrace v, v_1, w \rbrace} = \emptyset$ and 
$S_{\lbrace v, v_1, w \rbrace} = Q_{\lbrace v, v_1, w \rbrace}$ results.  
Now, we obtain: 
$\sum_{z \in W \Delta W_1} (\chi_Z (z) + \chi_{Z_1} (z)) = 
\sum_{z \in W \Delta W_1} \chi_{\Omega} (z) = \#_2 (W \Delta W_1) = 
\#_2 W + \#_2 W_1$. 
Since $\#_2 W = \#_2 W_1 = 1$, this sum vanishes. 
This completes the proof of the implication \eqref{WW1_implic}  
for $W, W_1 \in \mathcal{P}_{\mathrm{odd}} (\Omega)$. 

2) Now consider the case where both $W, W_1 \in \mathcal{P}_{\mathrm{even}} (\Omega)$. 
First look at the expression \eqref{Weven} corresponding to $i_{W}^{b} (\psi) \in B^{T_c}$. 
We have a conjunction over $U \in \mathcal{P}_{\mathrm{odd}} (\Omega)$ in the right-hand side. 
In particular, $U$ can be equal to $\pnt{v}$, where $v \in W$. 
Is it possible to have $\forall v \in W : S_{\pnt{v}} \not = \emptyset$? 
We claim that the answer is: no. 
Indeed, if $S_{\pnt{v}}$ is not empty, then we have 
$\sum_{z \in \pnt{v} \Delta W} (\psi (z) + \wt{\tau} (\pnt{v}, z)) = c (\#_4 W - 1) + 1$. 
Apply summation over $v \in W$ and infer that $\wt{\tau} (\pnt{v}, z) = 
\wh{\tau} (v, z)$ and $\wh{\tau} (v, z) + \wh{\tau} (z, w) = c (2) + 1$. 
This yields $(\# W - 1) b (\#_4 W) + (\#_4 W (\#_4 W - 1)/ 2) (c (2) + 1) = 0$. 
If $\#_4 W = 0$, then one obtains $b (0) = 0$, which is impossible, since we already have 
a condition $b (0) = 1$. 
If $\#_4 W = 2$, then $b (2) + c (2) + 1 = 0$, again contradicting the earlier 
assumption $b (2) + c (2) = 0$. 
Therefore, there always exists $v \in W$, such that $S_{\pnt{v}} = \emptyset$. 
Similarly, we may analyze the expression for $i_{W_1}^{b} (\psi_1) \in (M \backslash B)^{T_c}$, 
and conclude that there exists $v_1 \in W_1$, such that $S_{\pnt{v_1}} = W_{\pnt{v_1}}$.  
Start with the case $W_1 = W$. 
Is it possible to have $\psi (\cdot) = \psi_1 (\cdot)$? 
Suppose that $\psi$ and $\psi_1$ coincide. 
For every $v \in W$, if $S_{\pnt{v}} \not =  \emptyset$, then 
we have $\psi (v) = \sum_{z \in W} \wh{\tau} (v, z) + b (\#_4 W) + b (1) + c (\#_4 W - 1) + 1$. 
If $S_{\pnt{v}} =  \emptyset$, then $S_{\pnt{v}} \not = Q_{\pnt{v}}$, and this implies 
$\psi_1 (v) = \sum_{z \in W} \wh{\tau} (v, z) + b (\#_4 W) + b (1) + c (\#_4 W - 1) + 1$. 
But $\psi_1 (v) = \psi (v)$, so we have the same expression for $\psi (v)$ in all $v \in W$. 
For the same reasons as mentioned above, the condition $\sum_{v \in W} \psi (b) = b (\#_4 W)$ 
yields a contradiction. Therefore, $\psi (\cdot)$ and $\psi_1 (\cdot)$ cannot be equal, 
and the implication of the form \eqref{WW1_implic} for $W = W_1$ is established. 
Now assume, that $W_1 \not = W$. 
It is necessary to show, that $\sum_{z \in W \Delta W_1} (\psi (z) + \psi_1 (z)) = 
c (\#_4 (W \Delta W_1)) + 1$, whenever 
$i_{W}^{b} (\psi) \in B^{T_c}$ and $i_{W_1}^{b} (\psi_1) \in (M \backslash B)^{T_c}$ exist. 
Look at the expression \eqref{Weven} for $i_{W}^{b} (\psi) \in B^{T_c}$. 
Specialize $U$ to a one-point set $U = \pnt{u}$, $u \in \Omega$. 
If $S_{\pnt{u}} \not = \emptyset$, then the value of $\psi (u)$ is known. 
How to find the values of $\psi (\cdot)$ in other points of $\Omega$? 
Actually, we do not need to know the value of $\psi (z)$ for each $z \in \Omega$, but just 
the sum $\sum_{z \in W \Delta W_1} \psi (z)$. 

Let us establish an auxiliary fact first.  
We have $W \in \mathcal{P}_{\mathrm{even}} (\Omega)$. 
Take any $U \in \mathcal{P}_{\mathrm{odd}} (\Omega)$. 
Then $U \Delta W$ is a subset of $\Omega$, and, moreover, 
$U \Delta W \in \mathcal{P}_{\mathrm{odd}} (\Omega)$, since 
$\#_2 U \Delta W = \#_2 U + \#_2 W$. 
Suppose, that $S_{U} \not = \emptyset$. 
Is it possible to have $S_{U \Delta W} \not = \emptyset$ as well? 
Suppose, that it is. 
For $U', U'' \in \mathcal{P} (V)$, denote 
\begin{equation*}
g_{b, c} (U', U'') := 
b (\#_4 U') + b (\#_4 U'') + c (\#_4 (U' \Delta U'')) + 1. 
\end{equation*} 
One has: 
\begin{gather*} 
\sum_{z \in U} \psi (z) = \sum_{z \in W} \wt{\tau} (W, z) + 
g_{b, c} (U, W), \\
b (\#_4 W) + \sum_{z \in U} \psi (z) = \sum_{z \in W} \wt{\tau} (U \Delta W, z) + 
g_{b, c} (U \Delta W, W). 
\end{gather*}
Sum the two equations and regroup the terms:  
\begin{equation*} 
\sum_{v \in W} 
\big[ 
\wt{\tau} (U, v) + \wt{\tau} (U \Delta W, v) 
\big] = b (\#_4 W) + g_{b, c} (U, W) + g_{b, c} (U \Delta W, W). 
\end{equation*}
Since $\wt{\tau} (U, v) = \sum_{z \in U} \wh{\tau} (v, z) + g_{b, c} (U, \pnt{v})$, and, 
similarly, $\wt{\tau} (U \Delta W, v) = \sum_{z \in U \Delta W} \wh{\tau} (v, z) + 
g_{b, c} (U \Delta W, \pnt{v})$, we obtain:   
\begin{multline*} 
\wt{\tau} (U, v) + \wt{\tau} (U \Delta W, v) = 
\sum_{z \in W} \wh{\tau} (v, z) + b (\#_4 U) + b (\#_4 (U \Delta W)) + \\ + 
c (\#_4 (U \Delta \pnt{v})) + 
c (\#_4 (U \Delta W \Delta \pnt{v})). 
\end{multline*} 
It remains to sum over $v \in W$, and reduce the 
sum with $\wh{\tau} (v, z)$ on the right-hand side, taking into account that 
$\wh{\tau} (v, v) \equiv b (1)$, and $\wh{\tau} (v, z) + \wh{\tau} (z, v) = c (2) + 1$, $z \not = v$.   
Since $\# W$ is even, expressing $\# (U \Delta W)$ in terms of $\# U$, $\# W$, and $\# (U \cap W)$, yields: 
\begin{multline*} 
\frac{m (m - 1)}{2} \big( c (2) + 1 \big) +  
t \big\lbrace 
c (n - 1) + c (n + m - 2 t + 1) + \\
c (n + 1) + c (n + m - 2 t - 1) 
\big\rbrace 
+ c (n) + c (n + m - 2 t) = \\ = 
b (n) + b (m) + b (n + m - 2 t) , 
\end{multline*}
where $m := \#_4 W$, $n := \#_4 U$, and $t := \#_4 (U \cap W)$. 
This equation should be valid for generic $W$ and $U$. 
The value of $m$ can be $0$ or $2$, the value of $n$ can be $1$ or $3$, and 
the value of $t$ can be $0$, $1$, $2$, or $3$.  
In total this yields $2 \times 2 \times 4 = 16$ variants. 
A straightforward (Maple) computation shows, that 
each of the variants reduces to one of the following four equations: 
either $b (0) = 0$, or $1 + b (2) + c (2) = 0$, or 
$b (0) + \sum_{i = 1, 3} (b (i) + c (i)) = 0$, or 
$1 + \sum_{i = 1, 2, 3} (b (i) + c (i)) = 0$. 
Each of the four equations contradicts the already imposed assumptions on $b (\cdot)$ and $c (\cdot)$. 
Therefore, the following fact is established (recall, that $\# W$ is even): 
\begin{equation*} 
\forall U \in \mathcal{P}_{\mathrm{odd}} (\Omega) : 
( \text{$S_{U}$ or $S_{U \Delta W}$} ) = \emptyset. 
\end{equation*}
In a similar way (recall, that $\# W_1$ is also even), one obtains: 
\begin{equation*}
\forall U \in \mathcal{P}_{\mathrm{odd}} (\Omega) : 
S_{U} = Q_{U} \text{\, or \,} S_{U \Delta W_1} = Q_{U \Delta W_1}. 
\end{equation*} 
Since $W \not = W_1$, there exists $e \in W \Delta W_1$. 
Take such $e$. 
Observe, that $\# (\pnt{e} \Delta W)$ and $\# (\pnt{e} \Delta W_1)$ are odd. 
Specializing $U$ to $\pnt{e} \Delta W$, one obtains two facts: 
1) $S_{\pnt{e} \Delta W} = \emptyset$ or $S_{\pnt{e}} = \emptyset$; 
2) $S_{\pnt{e} \Delta W} = Q_{\pnt{e} \Delta W}$ or $S_{\pnt{e} \Delta W \Delta W_1} = 
Q_{\pnt{e} \Delta W \Delta W_1}$. 
Similarly, specializing $U$ to $\pnt{e} \Delta W_1$, one obtains two more facts: 
3) $S_{\pnt{e} \Delta W_1} = \emptyset$ or $S_{\pnt{e} \Delta W \Delta W_1} = \emptyset$; 
4) $S_{\pnt{e} \Delta W_1} = Q_{\pnt{e} \Delta W_1}$ or $S_{\pnt{e}} = Q_{\pnt{e}}$.   
Look at the set $S_{\pnt{e}}$. It is either empty, or non-empty. 
If $S_{\pnt{e}} = \emptyset$, then, due to the fourth fact, 
$S_{\pnt{e} \Delta W_1} = Q_{\pnt{e} \Delta W_1}$. 
This, together with the third fact, implies 
$S_{\pnt{e} \Delta W \Delta W_1} = \emptyset$. From the second fact:  
$S_{\pnt{e} \Delta W} = Q_{\pnt{e} \Delta W}$. 
Now consider the second possibility, $S_{\pnt{e}} \not = \emptyset$. 
The first fact then implies $S_{\pnt{e} \Delta W} = \emptyset$. 
Hence, due to the second fact, $S_{\pnt{e} \Delta W \Delta W_1} = 
Q_{\pnt{e} \Delta W \Delta W_1}$. From the third fact:  
$S_{\pnt{e} \Delta W_1} = \emptyset$. Then the fourth fact yields 
$S_{\pnt{e}} = Q_{\pnt{e}}$. 
Therefore, we have an alternative: either 
\begin{gather*} 
S_{\pnt{e}} = \emptyset, \quad 
S_{\pnt{e} \Delta W \Delta W_1} = \emptyset, \\
S_{\pnt{e} \Delta W} = Q_{\pnt{e} \Delta W}, \quad 
S_{\pnt{e} \Delta W_1} = Q_{\pnt{e} \Delta W_1},  
\end{gather*}
or 
\begin{gather*} 
S_{\pnt{e}} = Q_{\pnt{e}}, \quad 
S_{\pnt{e} \Delta W \Delta W_1} = Q_{\pnt{e} \Delta W \Delta W_1}, \\
S_{\pnt{e} \Delta W} = \emptyset, \quad 
S_{\pnt{e} \Delta W_1} = \emptyset.   
\end{gather*}
In both cases there is a way to compute the sums 
$\sum_{z \in W \Delta W_1} \psi (z)$ and $\sum_{z \in W \Delta W_1} \psi_1 (z)$. 
In the first case, the values of $\sum_{z \in \pnt{e} \Delta W \Delta W_1} \psi_1 (z)$ and $\psi_1 (e)$ are known. 
Their sum yields $\sum_{z \in W \Delta W_1} \psi_1 (z)$. 
The sum $\sum_{z \in W \Delta W_1} \psi (z)$ should be computed as the sum of 
$\sum_{z \in \pnt{e} \Delta W} \psi (z)$ and $\sum_{z \in \pnt{e} \Delta W_1} \psi (z)$. 
The second case is dual to the first one (the roles of $\psi$ and $\psi_1$ have to be interchanged). 
So we always know $\sum_{z \in W \Delta W_1} (\psi (z) + \psi_1 (z))$. 
It remains to compute this value, and then, using the assumptions about $b (\cdot)$ and $c (\cdot)$, 
verify that it reduces to $c (\#_4 (W \Delta W_1)) + 1$. This is done by a straightforward computation. 
Consider, for example, the first option. 
One has: 
\begin{gather*} 
\sum_{z \in \pnt{e} \Delta W} \psi (z) = 
\sum_{v \in W} \wt{\tau} (\pnt{e} \Delta W, v) + 
g_{b, c} (\pnt{e}, W), \\
\sum_{z \in \pnt{e} \Delta W_1} \psi (z) = 
\sum_{v \in W} \wt{\tau} (\pnt{e} \Delta W_1, v) + 
g_{b, c} (\pnt{e} \Delta W_1, W). 
\end{gather*}
This yields: 
\begin{multline*}
\sum_{z \in W \Delta W_1} \psi (z) = 
\sum_{v \in W} 
\big[ 
\wt{\tau} (\pnt{e} \Delta W, v) + 
\wt{\tau} (\pnt{e} \Delta W_1, v)
\big] + c (1) + \\ 
+ c (\#_4 (W \Delta W_1) - 1) 
+ b (\#_4 (\pnt{e} \Delta W)) + b (\#_4 (\pnt{e} \Delta W_1)). 
\end{multline*}
Expanding the definitions of $\wt{\tau} (\cdot, \cdot)$ in the square brackets, 
and then taking into account that $\# W$ is even, we obtain: 
\begin{multline*} 
\sum_{z \in W \Delta W_1} \psi (z) = 
\sum_{v \in W} \sum_{z \in W \Delta W_1} \wh{\tau} (v, z) + 
b (\#_4 (\pnt{e} \Delta W)) + \\ + 
b (\#_4 (\pnt{e} \Delta W_1)) + 
c (1) + c (\#_4 (W \Delta W_1) - 1) + \\ + 
\sum_{v \in W} \big[ 
c (\#_4 (\pnt{e} \Delta W \Delta \pnt{v})) + 
c (\#_4 (\pnt{e} \Delta W_1 \Delta \pnt{v})) 
\big]. 
\end{multline*}
A similar computation yields: 
\begin{multline*} 
\sum_{z \in W \Delta W_1} \psi_1 (z) = 
\sum_{v \in W_1} \sum_{z \in W \Delta W_1} \wh{\tau} (v, z) + 
c (\#_4 (\pnt{e} \Delta W)) + \\ + c (\#_4 (\pnt{e} \Delta W_1)) + 
b (1) + b (\#_4 (W \Delta W_1) - 1) + \\ + 
\sum_{v \in W_1} \big[ 
c (\#_4 (\pnt{e} \Delta \pnt{v})) + 
c (\#_4 (\pnt{e} \Delta W \Delta W_1 \Delta \pnt{v}))
\big].
\end{multline*}
Now, sum these equalities. 
On the right-hand side a sum of the form 
$\sum_{z, v \in W \Delta W_1} \wh{\tau} (v, z)$ appears; it is easily 
computed using $\wh{\tau} (v, v) \equiv b (1)$, and 
for $z \not = v$, $\wh{\tau} (v, z) + \wh{\tau} (z, v) = c (2) + 1$. 
Hence an expression for 
$\sum_{z \in W \Delta W_1} (\psi (z) + \psi_1 (z))$ in terms of $b (\cdot)$ and 
$c (\cdot)$ is obtained. 
On the other hand, we have to verify that it is equal to $c (\#_4 (W \Delta W_1)) + 1$. 
Denote $m := \#_4 W$, $m_1 := \#_4 W_1$, and $t := \#_4 (W \cap W_1)$. 
Equate the two expressions mentioned  
and simplify the result taking into account, that $\# W$ and $\# W_1$ are even. 
It is necessary to consider the cases, 
$e \in W \backslash W_1$ and $e \in W_1 \backslash W$, but in the end 
the result is the same: 
\begin{multline*} 
\frac{m + m_1 - 2 t}{2} \big[ c (2) + 1 \big] + 
c (m) + c (m_1) + c (m + 2) + c (m_1 + 2) + \\ + 
t \lbrace c (m_1) + c (m_1 + 2) 
+ c (m + m_1 - 2 t) + c (m + m_1 - 2 t + 2)
\rbrace + \\ + 
b (m - 1) + b (m_1 + 1) + b (1) + b (m + m_1 - 2 t - 1) + \\ +
c (m - 1) + c (m_1 + 1) + c (1) + c (m + m_1 - 2 t - 1) = \\ = 
c (m + m_1 - 2 t) + 1. 
\end{multline*}
It is straightforward to verify (best of all in Maple), that 
for all $m ,m_1 = 0, 2$ and all $t = 0, 1, 2, 3$, this equation reduces to one of the following: 
$1 + c (0) = 0$, of $\sum_{i = 1, 3} (b (i) + c (i)) = 0$, or 
$1 + c (0) + \sum_{i = 1, 3} (b (i) + c (i)) = 0$, or $0 = 0$. Due to the 
assumptions above, this always holds. Hence, it is established that 
if $W, W_1 \in \mathcal{P}_{\mathrm{even}} (\Omega)$, then  
any $i_{W}^{b} (\psi) \in B^{T_c}$ is in relation $T_c$ with any 
$i_{W_1}^{b} (\psi_1) \in (M \backslash B)^{T_c}$. 

3) Now consider the third possibility: let $W$ and $W_1$ both contain 
points outside $\Omega$, i.e. $W \cap \ov{\Omega} \not = \emptyset$, and 
$W_1 \cap \ov{\Omega} \not = \emptyset$. 
Take any $i_{W}^{b} (\psi) \in B^{T_c}$ and $i_{W_1}^{b} (\psi_1) \in (M \backslash B)^{T_c}$. 
Then we need to show, that $\psi_1 (\cdot) \not = \psi (\cdot)$. 
There is an expression \eqref{Wbeyond} for $\psi$ and the expression for $\psi_1$ is similar. 
For every $U \in \mathcal{P}_{\mathrm{odd}} (\Omega)$, such that 
$S_{U} \not = \emptyset$, 
the following quantity needs to be well defined: 
$\lambda_{U} := 
\sum_{z \in W \cap \ov{\Omega}} \varphi (z)$, 
where $\varphi$ is an element of $S_{U}$. 
Similarly, if $S_{U} \not = Q_{U}$, 
then the following quantity is well-defined:  
$\mu_{U} := 
\sum_{z \in W_1 \cap \ov{\Omega}} \varphi_1 (z)$, 
where $\varphi_1$ is an element of $Q_{U} \backslash S_{U}$. 
Let us start with the case $W_1 = W$. 
Note, that the set of values of $\sum_{z \in U \cap \ov{\Omega}} \phi(z)$ as 
$\phi$ varies over the entire $Q_{U}$ is $\mathbb{Z} / 2$.  
Therefore, if $S_{U} \not = \emptyset$, and $S_{U} \not = Q_{U}$, then 
one has $\mu_{U} = 1 + \lambda_{U}$. 
Is it possible to have $S_{U} = \emptyset$ or $S_{U} = Q_{U}$ at all? 
If $S_{U} = \emptyset$, then, in particular, $S_{U} \not = Q_{U}$, and, 
$\mu_{U}$ needs to be well-defined. At the same time, the corresponding sum 
$\sum_{z \in W_1 \cap \ov{\Omega}} \varphi_1$ 
ranges over $\mathbb{Z} / 2$ as $\varphi_1$ varies over 
$Q_{U} \backslash S_{U} = Q_{U}$. Hence, $\mu_{U}$ is not defined, and therefore, 
$S_{U}$ cannot be empty. 
For similar reasons, $S_{U}$ cannot be equal to $Q_{U}$. 
So, we have: $\forall U \in \mathcal{P}_{\mathrm{odd}} (\Omega) \, : \, S_{U} \not = \emptyset, Q_{U}$. 
If $W_1 = W$, we have to show, that $\psi (\cdot) \not = \psi_1 (\cdot)$. 
This follows from the fact, that  
$\sum_{z \in U} \psi (z) + \sum_{z \in W} \wt{\tau} (U, z) + g_{b, c} (U, W)$ should be equal to 
$\lambda_{U}$ and $\mu_{U} = 1 + \lambda_{U}$ at the same time, a contradiction! 
Now let $W_1 \not = W$. 
One needs to compute the sum $\sum_{z \in W \Delta W_1} (\psi (z) + \psi_1 (z))$. 
We can say nothing about the values of $\psi (z)$ and $\psi_1 (z)$ in the points 
$z \not \in \Omega$. Let us show, that these values are not needed, i.e. 
we show, that $W \Delta W_1 \subset \Omega$. 
The latter is equivalent to the statement, that the sets 
$K := W \cap \ov{\Omega}$ and $K_1 := W_{1} \cap \ov{\Omega}$ coincide. 
Indeed, for every $U \in \mathcal{P}_{\mathrm{odd}} (\Omega)$ the quantities 
$\lambda_{U} = \sum_{z \in K} \varphi (z)$ and 
$\mu_{U} = \sum_{z \in K_1} \varphi_1 (z)$ are defined ($\varphi \in S_U$, 
$\varphi_1 \in Q_{U} \backslash S_{U}$). 
The definition \eqref{QUset} of $Q_{U}$ implies, that the values of $\phi \in Q_{U}$ in the 
points outside $\Omega$ are not restricted by any condition. 
Hence, if $K \not = K_1$, there exists $\phi \in Q_{U}$, such that 
$\sum_{z \in K} \phi (z) = 1 + \lambda_{U}$ and 
$\sum_{z \in K_1} \phi (z) = 1 + \mu_{U}$. 
But such $\phi \not \in S_{U}, Q_{U} \backslash S_{U}$, a contradiction!  
Hence, $K = K_1$, and $W \Delta W_1 \subset \Omega$.  
Since $K = K_1$, it follows that $\mu_{U} = 1 + \lambda_{U}$, 
$U \in \mathcal{P}_{\mathrm{odd}} (\Omega)$. 
Take any $u \in \Omega$, and specialize $U$ to $\pnt{u}$. 
This yields: 
\begin{gather*} 
\psi (u) = \sum_{v \in W \cap \Omega} \wt{\tau} (\wh{u}, v) + \lambda_{\wh{u}} + 
g_{b, c} (\pnt{u}, W), \\
\psi_1 (u) = \sum_{v \in W_1 \cap \Omega} \wt{\tau} (\wh{u}, v) + (1 + \lambda_{\wh{u}}) + 
g_{b, c} (\pnt{u}, W_1). 
\end{gather*} 
Sum these two equalities, and then preform summation over 
$u \in W \Delta W_1$. The result should be $c (\#_4 (W \Delta W_1)) + 1$. 
Note, that on the other hand, the terms with $\wt{\tau} (\wh{u}, v)$ on the right-hand side 
are of the form $\sum_{u, v \in W \Delta W_1} \wt{\tau} (\wh{u}, v)$, and this 
sum can be expressed in terms of $b (\cdot)$ and $c (\cdot)$ as above. 
Denote $m := \#_4 W$, $m_1 := \#_4 W_1$, and $t := \#_4 (W \cap W_1)$. 
After simplifications, the result can be written in the form: 
\begin{multline*} 
 \frac{q (q - 1)}{2} 
\big[ c (2) + 1 \big] + 
q \big( 1 + b (1) + b (m) + b (m_1) \big) + \\ + 
(m - t) \big\lbrace c (m - 1) + c (m_1 + 1) \big\rbrace + \\ + 
(m_1 - t) \big\lbrace c (m + 1) + c (m_1 - 1) \big\rbrace + c (q) + 1 = 0, 
\end{multline*}  
where $q := m + m_1 - 2 t$. 
The variables $m$, $m_1$, and $t$, vary over $\mathbb{Z} / 4$. 
It remains to verify (easiest in Maple) 
that for each of the possible $4 \times 4 \times 4 = 64$ variants 
this equality is true. Each time the left-hand side reduces to one of the following variants: 
$1 + c (0)$, $1 + b (0)$, 
$b (2) + c (2)$,
$b (1) + b (3) + c (1) + c (3)$, 
or a linear combination of the mentioned ones. 
Hence, due to the imposed conditions, the equality is always valid. 
This means, that $(i_{W}^{b} (\psi), i_{W_1}^{b} (\psi_1)) \in T_c$ 
in case $W \cap \ov{\Omega} \not = \emptyset$ and 
$W_1 \cap \ov{\Omega} \not = \emptyset$.  

4) Now it is necessary to consider three mixed cases. 
Start with $W \in \mathcal{P}_{\mathrm{even}} (\Omega)$ and 
$W_1 \in \mathcal{P}_{\mathrm{odd}} (\Omega)$. 
Assume, that $i_{W}^{b} (\psi) \in B^{T_c}$ and 
$i_{W_1}^{b} (\psi_1) \in (M \backslash B)^{T_c}$. 
For $\psi_1$ there are two possibilities. 
The first one is that $\psi_1 \in S_{W}$, and hence 
$\psi_1 (z) = \wt{\tau} (W_1, z)$, $z \in \Omega$.  
The other is that $\psi_1 (\cdot)$ in the points $z \in \Omega$ 
is of the form: $\psi_1 (z) = \wt{\tau} (W_1, z) + \chi_Z (z)$, 
where $Z$ is some non-empty subset of $\Omega$, such that $\# (Z \cap W)$ is even. 
In the latter case, for all $U \in \mathcal{P}_{\mathrm{odd}} (\Omega)$  
such that $\# (U \cap Z)$ is odd, $S_{U} = Q_{U}$. 
Concerning $\psi (\cdot)$ one can say, that 
for all $U \in \mathcal{P}_{\mathrm{odd}} (\Omega)$, 
either $S_{U} = \emptyset$, or 
$\sum_{z \in U} \psi (z) = \sum_{z \in W} \wt{\tau} (U, z) + g_{b, c} (U, W)$. 
Consider the first possibility for $\psi_1$. In particular this 
implies that $S_{W_1} \not = \emptyset$. Hence the sum $\sum_{z \in W_1} \psi (z)$ 
is known. On the other hand, since $\psi_1 \in S_{W} \subset Q_{W}$, one has 
$\psi_1 (z \in \Omega) = \wt{\tau} (W_1, z)$. 
From this 
$\sum_{z \in W_1} \psi (z) = \sum_{z \in W} \psi_1 (z) + g_{b, c} (W, W_1)$, i.e. 
$(i_{W}^{b} (\psi), i_{W_1}^{b} (\psi_1)) \in T_c$ follows. 
Now consider the second possibility for $\psi_1$ (the one with $Z$). 
Observe, that $\#_2 (W \Delta W_1) = \#_2 W + \#_2 W_1 = 1$, i.e. 
$W \Delta W_1 \in \mathcal{P}_{\mathrm{odd}} (\Omega)$. 
Look at $(W \Delta W_1) \cap Z$. 
We have: $\#_2 (W \Delta W_1) \cap Z = \#_2 ((W \cap Z) \Delta (W_1 \cap Z)) = 
\#_2 (W \cap Z)$. 
Therefore, if $\# (W \cap Z)$ is odd, take $U = W \Delta W_1$ and obtain 
$S_{W \Delta W_1} = Q_{W \Delta W_1}$. In particular, $S_{W \Delta W_1} \not = \emptyset$, 
leading to the expression for $\sum_{z \in W \Delta W_1} \psi (z)$. 
The values of $\psi_1 (z)$ are known at all points $z \in \Omega$, so 
there is no problem to compute $\sum_{z \in W \Delta W_1} \psi_1 (z)$. 
Taking into account, that $\sum_{v \in W \Delta W_1} \chi_Z (v) = \#( (W \Delta W_1) \cap Z) = 1$, 
and then expressing $\wt{\tau}$ via $\wh{\tau}$ and $g_{b, c}$, 
we obtain: 
\begin{multline*} 
\sum_{z \in W \Delta W_1} \big[ 
\psi (z) + \psi_1 (z) 
\big] = 
1 + \sum_{v, z \in W} \wh{\tau} (v, z) + 
\sum_{v, z \in W_1} \wh{\tau} (v, z) + 
\\ 
\sum_{v \in W} g_{b, c} (\pnt{v}, W \Delta W_1) + 
\sum_{v \in W \Delta W_1} g_{b, c} (\pnt{v}, W_1) + 
g_{b, c} (W \Delta W_1, W). 
\end{multline*} 
On the other hand, this sum should be equal to $c (\#_4 (W \Delta W_1)) + 1$. 
Expressing the sums with $\wh{\tau} (v, z)$ in terms of $b (\cdot)$ and $c (\cdot)$, 
one obtains the following equality 
\begin{multline*} 
(m + m_1) b (1) + 
\Big( \frac{m (m - 1)}{2} + \frac{m_1 (m_1 - 1)}{2} \Big) \big[ c (2) + 1 \big] + \\
+ (m - t) c (q - 1) + t c (q + 1) + m \big\lbrace b (1) + b (q) + 1 \rbrace + \\ 
+ (m - t) c (m_1 + 1) + (m_1 - t) c (m_1 - 1) + \\
+ q \big\lbrace b (1) + b (m_1) + 1 \big\rbrace 
+ b (q) + b (m) + c (m_1) + c (q) + 1 = 0, 
\end{multline*}
where $m := \#_4 W$, $m_1 := \#_4 W_1$, $t := \#_4 (W \cap W_1)$, 
$q := m + m_1 - 2 t$. 
It is necessary to verify that this equality is true 
for every $m = 0, 2$, every $m_1 = 1, 3$, and $t = 0, 1, 2, 3$. 
This is done by a straightforward computation (in Maple). 
In each variant, the left-hand side 
reduces to a linear combination of the expressions 
$1 + b (0)$, $1 + c (0)$, $b (2) + c (2)$, and $\sum_{i = 1, 3} (b (i) + c (i))$. 
Due to the conditions on $b (\cdot)$ and $c (\cdot)$ imposed above, 
the equality is always true, so one 
has $(i_{W}^{b} (\psi), i_{W_1}^{b} (\psi_1)) \in T_c$ for 
$W \in \mathcal{P}_{\mathrm{even}} (\Omega)$, 
$W_1 \in \mathcal{P}_{\mathrm{odd}} (\Omega)$. 

5) Now consider the next case. Suppose that there exist  
$i_{W}^{b} (\psi) \in B^{T_c}$, 
$i_{W_1}^{b} (\psi_1) \in (M \backslash B)^{T_c}$, where  
$W_1 \cap \ov{\Omega} \not = \emptyset$ and 
$W \in \mathcal{P}_{\mathrm{even}} (\Omega)$. 
Take any $U \in \mathcal{P}_{\mathrm{odd}} (\Omega)$. 
For $\psi$ we have: 
$S_{U} = \emptyset$ or 
$\sum_{z \in U} \psi (z) = 
\sum_{v \in W} \wt{\tau} (U, v) + g_{b, c} (U, W)$. 
For $\psi_1$ we have: 
$S_{U} = Q_{U}$ or 
$\sum_{z \in U} \psi_1 (z) = 
\sum_{v \in W_1 \cap \Omega} \wt{\tau} (U, v) + 
\mu_{U} + g_{b, c} (U, W_1)$, where 
$\mu_U = \sum_{z \in W_{1} \cap \ov{\Omega}} \varphi (z)$, 
$\varphi$ being an element of $Q_{U} \backslash S_{U}$. 
Observe, that $S_{U}$ cannot be empty, since 
either $S_{U} = Q_{U}$, or $\mu_{U}$ is defined. 
Therefore, we always know the sum $\sum_{z \in U} \psi (z)$. 
In particular, $U$ can be of the form $\pnt{w}$, where $w \in W$, then  
$\psi (w) = \sum_{v \in W} \wt{\tau} (w, v) + b (\#_4 W) + b (1) + 
c (\#_4 (W \Delta \pnt{w})) + 1$. Sum over all $w \in W$, and 
use the fact that $\# W$ is even. This yields 
$\sum_{w \in W} \psi (w) = \sum_{v, w \in W} \wt{\tau} (w, v)$. 
On the other hand this sum should be equal to $b (\#_4 W)$. 
Hence one derives: 
$b (m) = (m (m - 1)/ 2) [c (2) + 1]$, $m := \#_4 W$. 
If $m = 0$, one obtains $b (0) = 0$, and if $m = 2$, one obtains 
$b (2) = c (2) + 1$. In both cases this contradicts the assumptions on $b (\cdot)$ and $c (\cdot)$. 
This means, that the pair $(i_{W}^{b} (\psi), i_{W_1}^{b} (\psi_1))$ cannot exist. 
 
6) It remains to investigate just the case where one of the sets $W$ or $W_1$ is an odd subset of $\Omega$, 
and the other contains at least one point outside $\Omega$. 
Let $W \in \mathcal{P}_{\mathrm{odd}} (\Omega)$, and $W_1 \cap \ov{\Omega} \not = \emptyset$. 
Suppose, that $i_{W}^{b} (\psi) \in B^{T_c}$ and 
$i_{W_1}^{b} (\psi_1) \in (M \backslash B)^{T_c}$. 
First look at the condition for $\psi_1$. 
For any $U \in \mathcal{P}_{\mathrm{odd}} (\Omega)$, the set $S_{U}$ cannot be empty, 
since one has either $S_{U} = Q_{U}$ or the quantity 
$\mu_{U} := \sum_{z \in W_1 \cap \ov{\Omega}} \varphi (z)$ needs to be defined 
($\varphi$ is an element of $Q_{U} \backslash S_{U}$; if $\varphi$ varies over 
the entire $Q_{U}$, the sum ranges over the entire $\mathbb{Z}/ 2$ and $\mu_{U}$ is undefined). 
Now look at the condition for $\psi$. 
First investigate the possibility $\psi (z) = \wt{\tau} (W, z) + \chi_{Z} (z)$, $z \in \Omega$, 
for some non-empty $Z \subset \Omega$, with $\# (Z \cap W)$ even. If $\# Z$ is odd, then take $U = Z$. 
This yields $S_{Z} = \emptyset$, contradicting the previous fact.  
If $\# Z$ is even, then since $\# W$ is odd, there always exist a point $e \in W \backslash Z$. 
(this is implied by the facts that $\# W$ is odd and $\# (W \cap Z)$ is even, and therefore 
$\# (W \backslash Z)$ is odd). Put $U = \pnt{e} \sqcup Z$. This yields 
$S_{\pnt{e} \sqcup Z} = \emptyset$, again a contradiction. 
Hence the only possibility that remains for $\psi$ is $\psi \in Q_{W} \backslash S_{W}$. 
For this case the values of $\psi (\cdot)$ are known in every point of $\Omega$. 
Since such $\psi$ is assumed to exist, $S_{W} \not = Q_{W}$. 
Now, put $U = W$ in the condition for $\psi_1$ (one can do it since 
$W \in \mathcal{P}_{\mathrm{odd}} (\Omega)$). 
This yields $\sum_{z \in W} \psi_1 (z) = 
\sum_{z \in W_{1} \cap \Omega} \wt{\tau} (W, z) + \mu_{U} + g_{b, c} (W, W_1)$. 
But $\wt{\tau} (W, z)$ is just the value of $\psi (z)$. Recall, that the definition of 
$\mu_{U}$ contains an arbitrary function $\varphi \in Q_{U} \backslash S_{U}$. 
Take $\varphi = \psi$. In the result, one obtains 
$\sum_{z \in W} \psi_1 (z) = \sum_{z \in W_1} \psi (z) + g_{b, c} (W, W_1)$, i.e. 
$(i_{W}^{b} (\psi), i_{W_1}^{b} (\psi_1)) \in T_c$. 
This completes the proof that $T_c$ satisfies the main condition \eqref{T_main}. 
Applying the described construction to the set $A_b$ and relation $T_c$, one obtains a coherent orthoalgebra. 
\qed

\section{Absense of bivaluations}

Recall, that we have made the following assumptions in order to construct an orthoalgebra: 
$N$ is divisible by $4$, 
$b (0) = 1$, $c (0) = 1$, $b (2) + c (2) = 0$, and $\sum_{i = 1, 3} (b (i) + c (i)) = 0$. 
Let us show that such orthoalgebra cannot admit bivaluations. 
Take $N + 1$ elements of $\mathrm{Max} (\mathcal{P}_{T_c} (A_b), \subset)$: 
$N$ elements $B_{v} := \lbrace i_{\pnt{v}}^{b} (\sigma) \rbrace_{\sigma \in L_{b} (\pnt{v})}$, $v \in V$, 
and an element $\wh{B} := \lbrace i_{V}^{b} (\pi) \rbrace_{\pi \in L_{b} (V)}$. 
Note, that $\wh{B}$ is transformed into $B_{v}$ if one applies 
$\wh{\theta}_{\pnt{v}}^{(\ov{a})} \, \wh{\theta}_{V}^{(\ov{a})} \, 
\wh{\theta}_{\pnt{v}}^{(\ov{a})}$ to each of its elements. 
Recall, that the ground set of our orthoalgebra is $\mathcal{P}^{T_c} (A_{b})$. 
Every singleton $\lbrace l \rbrace$, where 
$l \in \big( \bigsqcup_{v \in V} B_{v} \big) \sqcup \wh{B}$ is in this ground set, 
$\lbrace l \rbrace \in \mathcal{P}^{T_c} (A_{b})$. 
For every $v \in V$ the sum $\oplus_{l \in B_{v}} \pnt{l}$ is defined and equals $A_{b}$, 
i.e. the $\mathbf{1}$ of the orthoalgebra. 
Also, $\oplus_{l \in \wh{B}} \pnt{l} = \mathbf{1}$. 
Assume that there exists a bivaluation $f : X_{b, c} \to \mathbb{B}$, where 
$X_{b, c}$ denotes the constructed orthoalgebra. 
One has the following equalities in $\mathbb{B}$: 
$\oplus_{l \in B_{v}} f (\pnt{l})= \mathbf{1}$, $v \in V$, and 
$\oplus_{l \in \wh{B}} f (\pnt{l})= \mathbf{1}$. 
Since $\oplus$ in $\mathbb{B}$ is defined just in three cases, 
$\mathbf{0} \oplus \mathbf{0}$, $\mathbf{1} \oplus \mathbf{0}$, and $\mathbf{0} \oplus \mathbf{1}$, 
one derives two statements: 
1) $\forall v \in V \, \exists ! l \in B_{v} : f (l) = \mathbf{1}$; 
2) $\exists ! l \in \wh{B} : f (l) = \mathbf{1}$. 
Denote these uniquely defined elements by $l_{v} \in B_{v}$, $v \in V$, and 
$\wh{l} \in \wh{B}$, respectively. 
Any pair $(l, l')$, $l' \not = l$, of these elements cannot be in $T_c$. 
Indeed, then $l \oplus l'$ would have been defined. 
Applying to it $f$,  
$f (l) \oplus f (l') = \mathbf{1} \oplus \mathbf{1}$, follows a contradiction! 
Write $l_{v} = i_{\pnt{v}}^{b} (\sigma_v)$, $\sigma_v \in L_{b} (\pnt{v})$, and 
$\wh{l} = i_{V}^{b} (\wh{\pi})$, $\wh{\pi} \in L_b (V)$. 
The definition of $T_c$ yields, for any $v, v_1 \in V$, $v \not = v_1$, that: 
\begin{gather*} 
\sigma_{v} (v_1) + \sigma_{v_1} (v) = c (2), \\
\sum_{z \in V \backslash \pnt{v}} 
\big( \pi (z) + \sigma_v (z) \big) = c (3).  
\end{gather*}
Take the sum over $v \in V$ for the second equality. 
Using the definition of $L_b (V)$ one obtains 
\begin{equation*}
(N - 1) b (0) + 
\sum_{\substack{z, v \in V, \\ z \prec v}} (\sigma_{z} (v) + \sigma_v (z) ) = N c (3), 
\end{equation*}
where $\prec$ is any order on $V$. 
Now, using the first equality, and then the fact that $N$ is divisible by $4$, one arrives at 
$b (0) = 0$. This contradicts the assumption $b (0) = 1$. 
Therefore, a bivaluation of $X_{b, c}$ cannot exist.

\section{Isomorphic orthoalgebras} 

There are several options for the choice of $b (\cdot)$ and $c (\cdot)$ 
satisfying the conditions of the theorem. 
Let us derive a \emph{sufficient} condition for two orthoalgebras of the form $X_{b, c}$ to be isomorphic. 
Select any $(b, c)$ satisfying the conditions of the theorem, and any $(b', c')$ satisfying 
the same conditions. Construct $A_b := \bigsqcup_{U \in \mathcal{P} (\Omega)} L_b (U)$ and 
$A_{b'} := \bigsqcup_{U \in \mathcal{P} (\Omega)} L_{b'} (U)$, and define 
the relations $T_{c}$ and $T_{c'}$ on $A_{b}$ and $A_{b'}$, respectively. 
Denote by $i_{U}^{b} : L_b (U) \rightarrowtail A_b$ and 
$i_{U}^{b'} : L_{b'} (U) \rightarrowtail A_{b'}$ the canonical injections. 
Suppose that there exists a bijective map $\wh{t} : A_{b} \os{\sim}{\to} A_{b'}$, 
such that $(l, l_1) \in T_c$ implies $(\wh{t} (l), \wh{t} (l_1)) \in T_{c'}$. 
Then this map induces a bijection $\mathcal{P}^{T_c} (A_{b}) \os{\sim}{\to} 
\mathcal{P}^{T_{c'}} (A_{b'})$, which establishes an isomorphism of the 
orthoalgebras $X_{b, c}$ and $X_{b', c'}$. 
Let us try to construct such a map and investigate what kind of relations 
between $b$, $c$, $b'$, and $c'$, emerge. 

The map $\wh{t}$ is defined by a collection of bijections 
$\lbrace t_{U} \rbrace_{U \in \mathcal{P} (V)}$, where $t_{U} : L_b (U) \os{\sim}{\to} L_{b'} (U)$.  
Let us search for $t_{U}$ in the form: 
\begin{equation*} 
t_{U} (\psi) (v) = \psi (v) + \alpha_{U} (v), 
\end{equation*}
where $\alpha_{U} (v)$ are some $\mathbb{Z} / 2$-valued parameters, 
$U \in \mathcal{P} (V)$, $v \in V$. 
Denote 
\begin{gather*} 
\wt{b}_{U} := b (\#_4 U) + b' (\#_4 U), \\ 
\wt{c}_{U, U_1} := 
c (\#_4 (U \Delta U_1)) + c' (\#_4 (U \Delta U_1)) + 
\wt{b}_{U} + \wt{b}_{U}. 
\end{gather*}
The requirement that $\sum_{v \in U} \psi (v) = b (\#_4 U) \Rightarrow
\sum_{v \in U} t_{U} (\psi) (v) = b' (\#_4 U)$, yields 
for every $U \in \mathcal{P} (V)$ an equation on $\alpha_{U} (\cdot)$: 
\begin{equation*} 
\sum_{v \in U} \alpha_{U} (v) = \wt{b}_{U}. 
\end{equation*} 
Similarly, for every $U, U_1 \in \mathcal{P} (V)$, $U \not = U_1$, 
the requirement that 
for any $\psi \in L_b (U)$, $\psi_1 \in L_{b} (U_1)$, 
$\sum_{v \in U \Delta U_1} (\psi (v) + \psi_1 (v)) = c (\#_4 (U \Delta U_1)) + 1 
\Rightarrow 
\sum_{v \in U \Delta U_1} (t_{U} (\psi) (v) + t_{U} (\psi_1) (v)) = c' (\#_4 (U \Delta U_1)) + 1$, 
yields an equation: 
\begin{equation*} 
\sum_{v_1 \in U_1} \alpha_{U} (v_1) + 
\sum_{v \in U} \alpha_{U_1} (v) = \wt{c}_{U, U_1}. 
\end{equation*} 
These equations are similar to the equations \eqref{aa_cuus}, \eqref{aa_bus}, for $a_{S, U} (v)$ derived above. 
It is not difficult to solve them. First consider the case:  
$\# U = 1$ and $\# U_1 = 1$. From this it follows, that 
$\alpha_{\pnt{v}} (z)$ is a solution of the corresponding system iff it is of the form
$\alpha_{\pnt{v}} (z) = \alpha_{\pnt{v}}^{(\nu)} (z)$, where  
\begin{equation*} 
\alpha_{\pnt{v}}^{(\nu)} (z) := 
\begin{cases} 
\nu (zv), & \text{if $z \prec v$} \\
\wt{b}_{\pnt{v}}, & \text{if $z = v$} \\
\nu (zv) + \wt{c}_{\pnt{z}, \pnt{v}}, & \text{if $z \succ v$},   
\end{cases}
\end{equation*} 
where $\nu : \mathcal{E} (V) \to \mathbb{Z} / 2$ is an arbitrary function, 
and $\prec$ is some chosen and fixed order on $V$. 
The other $\alpha_{Q} (v)$, $\# Q \geqslant 2$, may be found from 
the specialization 
$U = \pnt{v}$, $U_1 = Q$, in the equation with $\wt{c}_{U, U_1}$. 
This yields $\alpha_{Q} (v) = \alpha_{Q}^{(\nu)} (v)$, where  
\begin{equation*} 
\alpha_{Q}^{(\nu)} (v) = \wt{c}_{\pnt{v}, Q} + \sum_{z \in Q} \alpha_{\pnt{v}}^{(\nu)} (z). 
\end{equation*} 
Now consider the equations with $\wt{b}_{U}$, corresponding to $U = Q$, $\# Q \geqslant 2$, and 
the equations with $\wt{c}_{U, U_1}$, corresponding to 
$U = Q$, $U_1 = Q_1$, with $\# Q, \# Q_1 \geqslant 2$. 
This leads to the following solvability conditions: 
\begin{gather} 
\wt{c}_{Q, Q_1} + \sum_{z \in Q} \wt{c}_{\pnt{z}, Q_1} + 
\sum_{z_1 \in Q_1} \wt{c}_{Q, \pnt{z_1}} + 
\sum_{\substack{z \in Q, \\ z_1 \in Q_1}} 
\wt{c}_{\pnt{z}, \pnt{z_1}} = 0, 
\label{c_cond_isom}
\\
\sum_{v \in Q} \wt{c}_{\pnt{v}, Q} + 
\sum_{\substack{v, z \in Q, \\ z \succ v}} 
\wt{c}_{\pnt{v}, \pnt{z}} = 
\wt{b}_{Q} + \sum_{v \in Q} \wt{b}_{\pnt{v}}.  
\label{b_cond_isom}
\end{gather}
Note that if one formally takes $Q$ or $Q_1$ of cardinality $1$, the 
corresponding equality trivializes. 
Note also that this system of equations becomes the system of equations \eqref{c_cond}, \eqref{b_cond}, 
for $c_{Q, Q_1}^{(S)}$ and $b_{Q}^{(S)}$ investigated above, if one formally replaces 
$\wt{c}_{Q, Q_1}$ with $c_{Q, Q_1}^{(S)}$, and $\wt{b}_{Q}$ with $b_{Q}^{(S)}$. 
From the definition of $\wt{b}_{U}$, it is clear that its value is determined by $\#_4 U$, 
i.e. $\wt{b}_{U} = \beta (\#_4 U)$, where $\beta : \mathbb{Z} / 4 \to \mathbb{Z} / 2$ is some function. 
Similarly, $\wt{c}_{U, U_1}$ can be written as 
$\wt{c}_{U, U_1} = \gamma (\#_4 (U \Delta U_1)) + \beta (\#_4 U) + \beta (\#_4 U_1)$, where 
$\gamma : \mathbb{Z} / 4 \to \mathbb{Z} / 2$ is some function. 
It is necessary to substitute these expressions into the solvability 
conditions \eqref{c_cond_isom}, \eqref{b_cond_isom}, above. 
The values of the resulting expressions are determined by $m := \#_4 Q$, $m_1 := \#_4 Q_1$, and 
$t := \#_4 (Q \cap Q_1)$. Analyzing the $4 \times 4 \times 4 = 64$ 
corresponding variants, we should discover which assumptions on $\beta (\cdot)$ and $\gamma (\cdot)$ emerge. 
The equation with four $\wt{c}$ yields: 
\begin{multline*} 
\big\lbrace \gamma (m + m_1 - 2 t) + \beta (m) + \beta (m_1) \big\rbrace + \\ + 
(m - t) \gamma (m_1 + 1) + t \gamma (m_1) + m \big[ \beta (1) + \beta (m_1) \big] + \\ +  
(m_1 - t) \gamma (m + 1) + t \gamma (m) + m_1 \big[ \beta (1) + \beta (m) \big] + \\ +
\Big( 
\frac{t (t - 1)}{2} + (m - t) t + (m_1 - t) t + (m - t) (m_1 - t)
\Big) \gamma (2) = 0. 
\end{multline*}
The equation with $\wt{b}$ yields 
\begin{equation*} 
m \lbrace \gamma (m - 1) + \beta (1) + \beta (m) \rbrace + 
\frac{m (m - 1)}{2} \gamma (2) = 
\beta (m) + m \beta (1). 
\end{equation*} 
A straightforward (Maple) computation shows, that 
all these equalities hold iff 
\begin{gather*} 
\gamma (0) = 0, \quad \gamma (2) = 0, \quad \gamma (1) + \gamma (3) = 0, \\
\beta (0) = 0, \quad \beta (2) = 0, \quad \beta (1) + \beta (3) = 0. 
\end{gather*} 
Recall that $b' (i) = b (i) + \beta (i)$, $c' (i) = c (i) + \gamma (i)$, $i \in \mathbb{Z} / 4$. 
Hence, if the additions $\beta (\cdot)$ and $\gamma (\cdot)$ satisfy these conditions, 
the corresponding orthoalgebras are isomorphic.

\section{The projective lines}

Let $N = 4$. 
Put $b (0) = c (0) = 1$ and the other $b (i) = c (i) = 0$, $i = 1, 2, 3$. 
We have the set $A_b$ and the relation $T_c \subset A_{b} \times A_{b}$. 
Let us write just $A$ and $T$ in this case. 
Let us show how (only in this particular case) the main condition on $T$ 
may be established in a different way (not combinatorial, but geometric).  

Take a Hilbert space $\mathcal{H}$ of finite dimension $d = 2^{N - 1} = 8$ over $\mathbb{C}$. 
Consider $\mathbb{P} (\mathcal{H})$ equipped with the orthogonality relation $\perp$. 
Suppose, that there exists an \emph{injective} map $\mu : A \rightarrowtail \mathbb{P} (\mathcal{H})$, 
such that $\forall x, x_1 \in A : (x, x_1) \in T \Leftrightarrow \mu (x) \perp \mu (x_1)$. 
Then the main property for $T$ can be easily established. 
Indeed, take any $M \in \mathrm{Max} (\mathcal{P}_{T} (A), \subset)$, and then 
any $B \subset M$. The map $\mu$ sends $M$ into a set of $d$ pairwise orthogonal 
projective lines. 
An element $x \in A$ falls into $B^{T}$ iff it's image $\mu (x)$ is orthogonal to 
every $\mu (y)$, $y \in B$. The latter is equivalent to 
$\mu (x) \in \big( \mathrm{span} \lbrace \mu (y) \, | \, y \in B \rbrace \big)^{\perp} =: P_1$. 
Similarly, $x \in A$ falls into $(M \backslash B)^{T}$ iff 
$\mu (x) \in \big( \mathrm{span} \lbrace \mu (y) \, | \, y \in 
M \backslash B \rbrace \big)^{\perp} =: P_2$. 
Since $\mu (y)$, $y \in M$, are pairwise orthogonal and the span over them is 
the whole space $\mathcal{H}$, 
the subspaces $P_1$ and $P_2$ have trivial intersection and are mutually orthogonal. 
For any $x_1 \in B$ and any $x_2 \in M \backslash B$, we have 
$\mu (x_1) \in P_1$ and $\mu (x_2) \in P_2$. Hence $\mu (x_2) \perp \mu (x_1)$, and this is equivalent to 
$(x_1, x_2) \in T$. So the main property of $T$ is established. 

In case $N = 4$ the map $\mu$ mentioned can be constructed. 
This fact relies on the results of \cite{RuugeFVO}. 
Put $\mathcal{H} = (\mathbb{C}^2)^{\otimes 3}$. 
Take any orthonormal basis $\lbrace \varphi_{\alpha} \rbrace_{\alpha}$ in $\mathbb{C}^2$ indexed 
by $\alpha \in \mathbb{Z} / 2$. 
Define a map $u : (\mathbb{Z} / 2)^2 \to \mathbb{R}$ as follows: 
$u (1, 1) := - 1$ and $u (i, j) := 1$ for $(i, j) \not = (1, 1)$. 
Construct another orthonormal basis $\lbrace \psi_{\beta} \rbrace_{\beta \in \mathbb{Z} / 2}$ in 
$\mathbb{C}^2$ by defining $\psi_{\beta} := (1 / \sqrt{2}) \sum_{\alpha} u (\alpha, \beta) \varphi_{\alpha}$. 
Recall that $A = \bigsqcup_{U \in \mathcal{P} (V)} L (U)$, 
where $L (U)$ consists of all functions $\phi : V \to \mathbb{Z} / 2$, such that 
$\sum_{z \in U} \phi (z) = b (\#_4 U)$. 
Note that since we have $b (0) = 1$, the set $L (\emptyset)$ is empty. 
Hence the latter disjoint union can be 
viewed as being taken over $U \in \mathcal{P} (V)^{\times} := 
\mathcal{P} (V) \backslash \lbrace \emptyset \rbrace$.  
Denote by $i_{U} : L (U) \rightarrowtail A$ the canonical injections. 
The elements $\mu ( i_{U} (\phi)) \in \mathbb{P} (\mathcal{H})$, 
$U \in \mathcal{P} (V)^{\times}$, $\phi \in L (U)$, are defined as follows.  
In \cite{RuugeFVO} there were defined  
120 projective lines in $\mathcal{H}$ denoted
by $\Psi_{\sigma}^{v}$, $X_{\varkappa}^{vw}$, $\Phi_{\rho}^{v}$, and $F_{\pi}$, where 
$v, w \in V$, $w \not = v$, and the indices $\sigma$, $\varkappa$, $\rho$ and $\pi$ 
vary over the sets $S_{v}$, $K_{vw}$, $R_{v}$ and $\Lambda$, respectively, 
defined as follows: 
\begin{gather*} 
S_{v} := \mathrm{Maps} (
\lbrace \lbrace v, z \rbrace \, | \, z \in V \backslash \lbrace v\rbrace \rbrace
 \to \mathbb{Z} / 2), \\
K_{vw} := \mathrm{Maps} (\lbrace \lbrace v, w \rbrace \rbrace \sqcup 
(V \backslash \lbrace v, w \rbrace) \to \mathbb{Z} / 2), \\
R_{v} := \mathrm{Maps} (\lbrace \lbrace z, w \rbrace \, | \, z, w \in 
V \backslash \lbrace v \rbrace, z \not = w\rbrace \to \mathbb{Z} / 2), \\
\Lambda := \big\lbrace 
\pi \in \mathrm{Maps} (V \to \mathbb{Z} / 2) \, \big| \, 
\sum_{z \in V} \pi (z) = 1 \big\rbrace. 
\end{gather*} 
The corresponding formulae for the projective lines are given 
in terms of $\varphi_{\alpha}$, 
$\psi_{\beta}$, and $u (\alpha, \beta)$ ($\alpha, \beta \in \mathbb{Z}/ 2$),  
and discussed in more detail in that paper. 
Note that the index sets $S_v$, $K_{vw}$, $R_v$ and $\Lambda$, all have 
cardinality $2^{3} = 8$. 

Let us establish bijections 
$\alpha_{v} : L (\pnt{v}) \os{\sim}{\to} S_{v}$, 
$\beta_{vw} : L (\pnt{v, w}) \os{\sim}{\to} K_{vw}$, 
$\gamma_{v} : L (V \backslash \lbrace v \rbrace) \os{\sim}{\to} R_{v}$, and 
$\lambda : L (V) \os{\sim}{\to} \Lambda$. 
For $v \in V$ and $\phi \in L (\pnt{v})$, put:  
\begin{equation*} 
\alpha_{v} (\phi) (vz) := \phi (z), \quad z \in V \backslash \pnt{v}. 
\end{equation*}
For $v, w \in V$, $w \not = v$, and $\phi \in L (\lbrace v, w \rbrace)$, put 
\begin{gather*} 
\beta_{vw} (\phi) (vw) \, := \, \phi (v) = \phi (w), \\
\beta_{vw} (\phi) (z) := \phi (t), \quad 
\beta_{vw} (\phi) (t) := \phi (z), 
\end{gather*}
where $z$ and $t$ are the two different elements of $V \backslash \lbrace v, w \rbrace$.  
(Note, that $\phi (v) = \phi (w)$, since $b (2) = 0$.)
For every $v \in V$ and $\phi \in L (V \backslash \pnt{v})$, put 
\begin{equation*} 
\gamma_{v} (\phi) (V \backslash \lbrace v, z \rbrace) := 
1 + \phi (v) + \phi (z), \quad z \in V \backslash \lbrace v \rbrace. 
\end{equation*}
Finally, for every $\phi \in L (V)$, put 
\begin{equation*}
\lambda (\phi) (v) := 1 + \phi (v), \quad v \in V. 
\end{equation*}

The collection of bijections $\alpha_{v}$, $\beta_{vw}$, $\gamma_{v}$, and $\lambda$, 
define an injective map $\mu : A \rightarrowtail \mathbb{P} (\mathcal{H})$ by the formulae  
$i_{\pnt{v}} (\phi_1) \mapsto \Psi_{\alpha_v (\phi_1)}^{v}$, 
$i_{\lbrace v, w \rbrace} (\phi_2) \mapsto X_{\beta_{vw} (\phi_2)}^{vw}$, 
$i_{V \backslash \pnt{v}} (\phi_3) \mapsto \Phi_{\gamma_{v} (\phi_3)}^{v}$, and 
$i_{V} (\phi_0) \mapsto F_{\lambda (\phi_0)}$, where 
$\phi_1 \in L (\pnt{v})$, 
$\phi_2 \in L (\lbrace v, w \rbrace)$, 
$\phi_3 \in L (V \backslash \pnt{v})$, and 
$\phi_0 \in L (V)$. 
It is straightforward to verify that it transforms a pair $(i_{U} (\phi), i_{W} (\phi')) \in T$ 
($\phi \in L (U)$, $\phi' \in L (W)$, $U, W \in \mathcal{P} (V)^{\times}$), into a 
pair of orthogonal projective lines. 
Note, that the only facts needed in order to prove this, are the following four properties of 
$u (\cdot, \cdot)$: 
1) $u (\alpha, \beta) = u (\beta, \alpha)$; 
2) $u (\alpha, \beta + \gamma) = u (\alpha, \beta) u (\alpha, \gamma)$; 
3) $\sum_{\beta \in \mathbb{Z} / 2} u (\alpha, \beta) u (\beta, \alpha) = 2 \delta_{\alpha, \gamma}$; 
4) $u (\alpha, 1 + \alpha) \equiv 1$. 
Therefore $T$ satisfies the main condition. 

Let us mention, how to obtain in principle the formulae for the projective lines 
(for $N = 4$). 
The result will be just the 120 projective lines constructed in \cite{RuugeFVO}. 
The configuration of these lines is saturated (i.e. every subset of pairwise orthogonal lines 
is contained in a set of eight pairwise orthogonal lines) and 
has a Kochen-Specker-type property \cite{KochenSpecker}. 
More precisely, this set contains a subset of $40 = 5 \times 8$ projective lines, which 
are implicitly present in the no-hidden-variables argument due to D. Mermin \cite{Mermin}. 
Denote the lines as $l_{\phi}^{U}$, $U \in \mathcal{P} (V)^{\times}$, $\phi \in L (U)$. 
It is convenient to view the set of four points $V$ as 
a disjoint union of the ground set of $\mathbb{Z} / 3$ and a singleton $\lbrace \ast \rbrace$, 
where $\ast$ is a formal symbol. 
Write $\mathbb{Z} / 3$ additively, and denote its elements as $0$, $1$, and $2$.  
For $\xi \in L (\pnt{\ast})$, put 
\begin{equation*}
l_{\xi}^{\pnt{\ast}} := 
\mathbb{C} \bigotimes_{i \in \mathbb{Z} / 3} \os{2 i}{\varphi}_{\xi (i)}, 
\end{equation*} 
where the upper indices denote the ordering of the factors in the tensor product.
Let $\lbrace \varphi_{\alpha} \rbrace_{\alpha}$ and 
$\lbrace \psi_{\beta} \rbrace_{\beta}$ be the orthonormal bases in $\mathbb{C}^2$ as above. 
For $k \in \mathbb{Z} / 3$ and $\xi \in L (\pnt{k})$, put 
\begin{equation*} 
l_{\xi}^{\pnt{k}} := \mathbb{C} 
\big\lbrace 
\os{2 k}{\varphi}_{\xi (\ast)} \otimes 
\bigotimes_{j \in (\mathbb{Z} / 3) \backslash \lbrace k \rbrace}
\os{k + j}{\psi}_{\xi (j)}
\big\rbrace. 
\end{equation*}
For any $U \in \mathcal{P} (V)$, $\# U \geqslant 2$, 
one can search for the projective line corresponding to $\eta \in L (U)$ in the form 
\begin{equation*} 
l_{\eta}^{U} := \sum_{\xi \in L (\pnt{\ast})} 
B_{\eta}^{U} (\xi) \, 
\bigotimes_{i \in \mathbb{Z} / 3} \os{2 i}{\varphi}_{\xi(i)}, 
\end{equation*}
where $B_{\eta}^{U} (\xi) \in \mathbb{C}$ are some coefficients. 
The conditions 
\begin{equation*} 
\sum_{z \in U \Delta \pnt{v}} \big( \eta (z) + \xi (z) \big) = 
c (\#_4 (U \Delta \pnt{v})) + 1 
\Rightarrow 
l_{\xi}^{\pnt{v}} \perp l_{\eta}^{U}, 
\end{equation*}
where $v$ varies over $V$, and $\xi$ varies over $L (\pnt{v})$, 
yield (for every $U$ and $\eta$) a system of equations on $\lbrace B_{\eta}^{U} (\xi) \rbrace_{\xi}$. 
This system is homogeneous and linear, but overdetermined. 
Nevertheless, it turns out that it has non-trivial solutions. 
Moreover, for every $U \in \mathcal{P} (V)$, the obtained projective lines 
$\lbrace l_{\eta}^{U} \rbrace_{\eta \in L (U)}$ are pairwise orthogonal, 
and for every $U, U_1 \in \mathcal{P} (V)$, 
if $U \not = U_1$, then $l_{\eta}^{U} \not = l_{\eta_1}^{U_1}$, where  
$\eta \in L (U)$, $\eta_1 \in L (U_1)$.  
A straightforward computation shows, that 
the orthogonality relation between the lines 
corresponding to different 
$U, U_1 \in \mathcal{P} (V)^{\times}$, $U \not = U_1$, 
is described by the formula 
\begin{equation*} 
l_{\eta}^{Q} \perp l_{\eta_1}^{Q_1} \quad  
\Leftrightarrow \quad 
\sum_{z \in Q \Delta Q_1} \big( 
\eta (z) + \eta (z_1)
\big) = c (\#_4 (Q \Delta Q_1)) + 1,  
\end{equation*}
where $\eta \in L (U)$, $\eta_1 \in L (U_1)$. 
It remains to define the injection $\mu : A \rightarrowtail \mathbb{P} (\mathcal{H})$ 
by the formula: 
$\mu (i_{U} (\phi)) := l_{\phi}^{U}$, $U \in \mathcal{P} (V)^{\times}$, $\phi \in L (U)$. 


\vspace{0.3 true cm}
The present work has been supported by the Liegrits programme of the European Science Foundation.



\begin{thebibliography}{99} 
\morespace


\bibitem{BirkhoffNeumann} 
Birkhoff, G.; von Neumann, J.,  
``The logic of quantum mechanics'', 
\emph{Ann. of Math.} \textbf{37}, no.\,4, 823 -- 843 (1936).  

\bibitem{DvurPulm} Dvure\v{c}enskij, A.; Pulmannov\'{a}, S. (2002).  
\emph{New Trends in Quantum Structures}.  
Mathematics and its Applications, 516.  
Kluwer Academic Publishers, Dordrecht; 
Ister Science, Bratislava, 2000. xvi + 541 pp.

\bibitem{FoulisBennett} 
Foulis, D.J.; Bennett, M.K., 
``Effect algebras and unsharp quantum logics'', 
Special issue dedicated to Constantin Piron on the occasion of his 
sixtieth birthday, 
\emph{Found. Phys.} \textbf{24}, no.\,10, 1331 -- 1352 (1994). 

\bibitem{FoulisGreechieRuttimann} 
Foulis, D.J.; Greechie, R.J.; R\"{u}ttimann, G.T., 
``Filters and supports in orthoalgebras'', 
\emph{Internat. J. Theoret. Phys.} \textbf{31}, no.\,5, 789 -- 807 (1992).    

\bibitem{Isham} Isham, C.J., 
``Topos theory and consistent histories: the internal logic of the set of all consistent sets'', 
\emph{Internat. J. Theor. Phys.} \textbf{36}, no.\,4, 785 -- 814 (1997). 

\bibitem{KochenSpecker} Kochen, S.; Specker, E.P., 
``The problem of hidden variables in quantum mechanics'', 
\emph{J. Math. and Mech.} \textbf{17}, 59 -- 87 (1967).    

\bibitem{Mackey} Mackey, G.W. (1963). 
\emph{Mathematical Foundations of Quantum Mechanics}.  
Benjamin, New York. 

\bibitem{Mermin} Mermin, D. 
``Hidden variables and the two theorems of John Bell'', 
\emph{Rev. Mod. Phys.} \textbf{65}, no.\,3, part 1, 803 -- 815 (1993). 

\bibitem{Neumann} Von Neumann, J. (1955).  
\emph{Mathematical Foundations of Quantum Mechanics}. 
Princeton University Press, Princeton, N.J.

\bibitem{Ruuge} Ruuge, A.E., ``Indeterministic objects in the category of effect algebras 
and the passage to the semiclassical limit'', 
\emph{Internat. J. Theor. Phys.} \textbf{43}, no.\,12, 2325 -- 2354 (2004).   

\bibitem{RuugeFVO} Ruuge, A.E.; Van Oystaeyen, F., 
``Saturated Kochen-Specker-type configuration of 120 projective lines in 
eight-dimensional space and its group of symmetry'', 
\emph{J. Math. Phys.} \textbf{46}, no.\,5, 052109, 28\,pp (2005).



\end{thebibliography}
\end{document}